\documentclass[preprint,12pt,authoryear]{elsarticle}

\usepackage{multicol}
\usepackage{color}
\usepackage{xspace}
\usepackage{amsmath,amssymb,mathtools}
\usepackage{enumerate}

\usepackage{graphicx}
\usepackage{dcolumn}
\usepackage{bm}
\usepackage{ulem}
\newcommand\simabove{\underset{\mathclap{\scriptsize\mbox{$\theta \gg \theta_\mathrm{c}$}}}{\simeq}}
\newcommand\simbelow{\underset{\mathclap{\scriptsize\mbox{$\theta \ll \theta_\mathrm{c}$}}}{\simeq}}

\newcommand{\lambdab}{$\lambda_\mathrm{b}~$}

\title{From bending to stretching driven peeling of heterogeneous adhesives}
\author{Laurent Ponson}
\ead{laurent.ponson@sorbonne-universite.fr}
\affiliation{organization={Institut Jean le Rond d'Alembert (UMR 7190), CNRS and Sorbonne Université},
city={Paris},
country={France}}

\begin{document}

\begin{abstract}
We study theoretically the peeling behavior of adhesives. Adopting a fracture mechanics approach, we derive the equation of motion of the adhesion front propagating at the interface between the adhesive and the substrate from which the peel strength is inferred. The originality of our approach lies in the description of the interplay during peeling between the stretching and the bending modes of deformation of the adhesive that is described as a F{\"o}ppl-Von Karman's thin film. Considering first a straight adhesion front, we retrieve the most salient feature of homogeneous adhesives, namely a peeling angle dependent peel strength driven by bending at large angles and by stretching at low angles. We also derive the shape of the adhesive that can be described using a single bending length scale derived from our model. We then investigate the impact of adhesion heterogeneities. We evidence that the deformations of the adhesion front are governed by a non-local interface elasticity the strength of which decreases with the peeling angle. This phenomenon reflects the transition between a stretching dominated peeling at low angle to a bending driven peeling at large angles that is captured in our model. This transition impacts the stability of adhesive fronts that relaxe more slowly from perturbations and gives rise to a stronger toughening effect in presence of a disorder distribution of adhesion energy at low peeling angles. Overall, this study sheds light on the central role played the elastic deformations of adhesives on their peeling behavior. The proposed framework unfolds the complex interplay between the deformation of adhesives and the peeling driving force that may be leveraged to engineer heterogeneous adhesives with enhanced properties. It also provides rich insights on the mechanisms underlying the emergence of non-local elasticity in interface problems.
\end{abstract}

\maketitle

\vspace{20 pt}
Moving interfaces are ubiquitous in solids. They govern the ability of materials to sustain mechanical loads~\citep{Lawn}. They also drive their magnetic properties~\citep{Landeau2}, their wettability~\citep{DeGennes2} or their phase transformation~\citep{Fultz}. And they can give rise to unexpected and highly complex material behaviors such as superconductivity~\citep{Combescot} and superelasticity in shape memory alloys~\citep{Bhattacharya2}. The ability of continuum mechanics to capture and predict the rich behavior of moving interfaces appears as one of its most glorious successes.

The cornerstone in interface modeling is the derivation from thermodynamics of a driving force that promotes its motion and a resistance that opposes to it. Such an approach encapsulates the rich behavior of interfaces into an equation of motion that describes them as driven manifolds. A striking illustration of the success of interface modeling is linear elastic fracture mechanics. Fracture in solids results from the cooperative dynamics of damage processes as diverse as coalescence of ductile cavities, micro-cracking, and crazing. Yet, such a diversity can not only be encrypted in a rather simple interface equation of motion, but this equation is essentially the same for all elastic materials~\citep{Gao, Schmittbuhl4, Bonamy5,Ponson23}. 

However, modeling interfaces at a continuum level though a coarse-grained equation of motion is not without raising challenges and its own degree of complexity. The most delicate aspect of this approach is the introduction of an interface elasticity that describes the interface response as it deforms. It turns out that this emerging elasticity is very often {\it non-local}: A localized advance somewhere along the interface may trigger the advance of the interface somewhere else. When interfaces are governed by the bulk elasticity of the surrounding solid, like e.g. in fracture, the interface elasticity is {\it fully} non-local: The interactions are shown to decay as $1/r^2$ with the distance $r$ between two regions of the interface, and the localized motion of the interface impacts the response of the interface everywhere else~\citep{Rice4}. These fully non-local interactions constitute the very essence of their complex behavior. They give rise to the rich phenomenology observed in fracture experiments, such as the trapping and untrapping of cracks from tough obstacles~\citep{Dalmas2,Chopin5} or their collective pinning and depinning in presence of disorder~\citep{Schmittbuhl4,Bonamy5,Chopin3}. They also underlie the fingering instability that takes place in highly heterogeneous solids~\citep{Vasoya4}, More generally, crack front elasticity controls how microstructural heterogeneities affect the propagation of cracks. And thus, it largely impacts the {\it effective toughness} of heterogeneous brittle solids~\citep{Demery, Patinet2,Lebihain}

Following the pioneering work of~\cite{Rice4}, many theoretical studies have been dedicated to the derivation of the crack front elasticity for various configurations, including penny-shaped cracks~\citep{Gao3}, tunnel cracks~\citep{Leblond6}, coplanar slit-cracks~\citep{Pindra2,Legrand3} and interfacial cracks~\citep{Lazarus5}, under tensile loading conditions, but also under shear~\citep{Gao4,Lazarus4}, a line of research that has recently been reviewed by~\cite{Lazarus2}. Beyond linear long-range elasticity, large front deformations give rise to {\it non-linear} and non-local elasticity that has been studied for semi-infinite tensile cracks~\citep{Leblond2} and subsequently investigated experimentally~\citep{Patinet,Vasoya2}. 

\begin{figure}[h]
\includegraphics[width=12.cm]{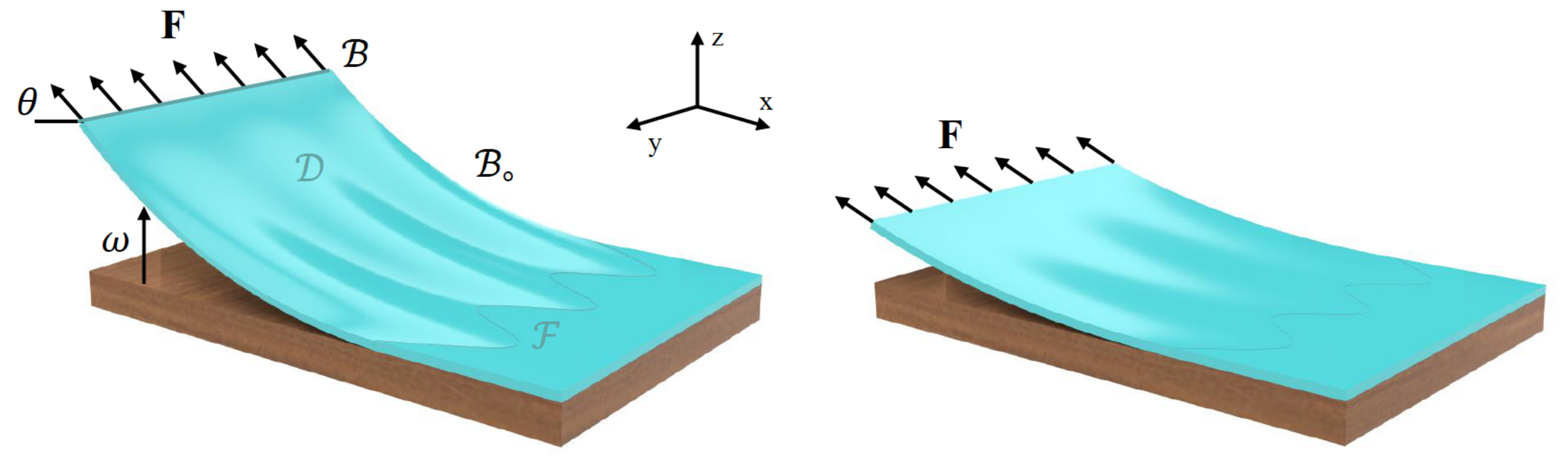}
\centering
\caption{Peeling of thin film adhesives in presence of adhesion heterogeneities at two different peeling angles $\theta$. The wrinkles that form at large peeling angle, but vanish at low peeling angle, are responsible for the non-local stiffness of the adhesion front~\citep{Xia3}. What is the front elasticity at low peeling angle? And how does it affect the overall peeling behavior of the adhesive? }
\label{Fig1}
\end{figure}

In parallel to these last developments, and nearly 30 years after their beginning, a similar approach has been proposed to describe the peeling behavior of thin film adhesives detaching from rigid substrates~\citep{Xia,Xia3}, see Fig.~\ref{Fig1}. In these works, the authors study how an inhomogeneous spatial distribution of adhesion energy at the substrate-adhesive interface generates deformations of the peeling front and ultimately impacts the peel strength of thin film adhesives. \cite{Xia} showed that the peeling front elasticity is fully non-local, as for crack fronts. Yet the {\it strength} of this interface elasticity is larger than the one of crack fronts. In other words, adhesion fronts are more reluctant to deform than cracks. This emergent stiffness stems from the deformation of the adhesive that accompany front perturbations: Perturbing the adhesion front results in {\it wrinkles} visible in Fig.~\ref{Fig1} that increase the bending energy of the film, maintaining the adhesion front as straight as possible.

Despite its success to capture the experimental properties of heterogeneous adhesives at {\it large} peeling angles~\citep{Xia,Xia3}, this mechanism appears to be unsuitable to describe their low angle peeling response. Indeed, in the limit case of a pure shear peeling at zero angle, the out-of-plane deformations of the adhesive and thus the wrinkles vanish. Does this imply that peeling fronts can deform without any resistance at low peeling angles ?

In this study, we address this question theoretically. We explore how the adhesion front elasticity, and thus the peeling behavior, change as the peeling angle is varied. We describe the adhesive as a F{\"o}ppl-Von Karman thin film so that we not only take into account the deformations resulting from its {\it bending} as in~\cite{Xia,Xia3}, but also from its {\it stretching}. As a result, the proposed model captures the {\it interplay} between both modes of deformation. It first provides a unified framework for predicting the strength and the shape of {\it homogeneous} adhesives that goes beyond the models currently available~\citep{Rivlin,Kendall}. It then reveals that the adhesion fronts are subjected to a {\it peeling angle dependent elasticity}. In other words, the ability of adhesion fronts to deform depends on the peeling angle, thus impacting the peel strength in presence of heterogeneous adhesion energy.

Beyond the effective strength of heterogeneous adhesives, our analysis sheds light on the {\it stability} of adhesion fronts and its dependance to the peeling angle. Adhesion fronts can be destabilized and form fingers~\citep{Ghatak}, changing radically the peeling resistance of thin film adhesives~\citep{Shen_J}, an instability particularly prominent in presence of shear at low peeling angle~\citep{Sosson}. In the following, we determine whether the adhesion front elasticity remains non-local at low peeling angles and opposes to front deformations. On a broader perspective, we investigate here how interfaces behave when their elastic bulk counterpart – here, the elastic thin film – admits two competing deformation modes – here, bending and stretching. We will show that in such a scenario, the front elasticity can crossover from one regime governed by one deformation mode to another governed by the second deformation mode. Our analysis will provide how and when adhesion fronts switch from one regime to another, and how this transition impacts the peeling behavior of adhesives.

Our paper is organized as follows. In Section~\ref{Sec1}, we use a variational approach to derive the equation of motion of the adhesion front of a peeled adhesive, that we specify to a F{\"o}ppl-Von Karman thin film. In Section~\ref{Sec2}, this equation of motion is used to study the response of an adhesive peeled from a {\it homogeneous} substrate, a 2D situation where the adhesion front remains straight. We derive the {\it peel strength} and the {\it shape} of the adhesive, two key features for interpreting experimental peel tests. We show that our analysis go beyond the predictions of other 2D peeling models~\citep{Rivlin,Kendall}. Section~\ref{Sec3} addresses the full 3D problem by considering a slightly perturbed adhesion front. We start, in Section~\ref{Sec_G}, by deriving the general expression of the driving force. As shown in Section~\ref{Sec_Decouple}, when small wavelength perturbations are considered, the driving force decomposes in two decoupled contributions, one associated to the stretching mode of deformation of the adhesive and another one associated to its bending mode. Section~\ref{Sec_Gs} derives the stress field in the deformed adhesive from which results the stretching contribution. The bending contribution resulting from the out-of-plane displacement of the adhesive is derived in Section~\ref{Sec_Gb}. Section~\ref{Sec_Wrinkle} explores the wrinkles that form on the thin film and discusses the mechanisms underlying the emergence of long-range interactions as well as their cut-off at large perturbation wavelengths. The main results are compiled and discussed in Section~\ref{Sec4}. Section~\ref{Sec_thetac} provides the peeling angle dependent front elasticity and introduces a critical peeling angle delimiting stretching from bending driven peeling. The implications of our findings on the stability of adhesion fronts and on the effective peel resistance in presence of adhesion heterogeneities are investigated in Section~\ref{Sec_Stability}. Finally, Section~\ref{Sec5} summarizes our main findings, presents some promising perspectives and discusses the implications of our work for the design of adhesives with new and enhanced peeling properties.

\section{General expression of the peeling driving force}
\label{Sec1}
We study theoretically the peeling of an elastic thin film of thickness $h$ from a rigid substrate, as represented in Fig. \ref{Fig1}. We adopt a linear fracture mechanics perspective, namely that all the non-elastic processes are confined along the peeling front in a process zone of size $\ell_\mathrm{c} \ll h$.

We note $\mathcal{D}$ the domain of the film already peeled from the substrate and $\mathcal{F}$ the adhesion front separating the bonded region of the film from the peeled one. Our objective is to describe the behavior of the adhesion front as it deforms, {\it e.g.} under the effect of adhesion heterogeneities. The adhesion front parallel to the $y$-axis propagates at the interface between the film and the substrate along the $x$-axis. Indicating vectors by bold face, $ {\bf x} = \{ x \, y \} $ provides the position at this interface and parametrizes the position on the film in its reference configuration, {\it i.e.} before peeling. Together with the $z$-axis perpendicular to the film-substrate interface, they constitute the coordinate system used subsequently. With these notations, the peeled region of the adhesive extends in the semi-half space $x<0$. The adhesion front is located {\it on average} in $x=0$. The in-plane and out-of-plane displacements of the film are noted ${\bf u} = \{ u_\mathrm{x}({\bf  x}), u_\mathrm{y}({\bf x}) \}$ and $w(\bf{x})$, respectively. Debonding results from the application of an external force ${\bf F}$ defined by unit length along the $y$-axis with in-plane and out-of-plane components $F_x$ and $F_z$. ${\bf F}$ is applied in $\mathcal{B}$ at the extremity of the film with an angle $\theta = \arctan(F_z/F_x)$ with respect to the horizontal plane. The film lateral boundaries are noted  $\mathcal{B}_\circ$ and they remain free.

The elastic energy release rate drives the motion of the peeling front. Its distribution along the front that we seek to determine governs the front evolution and the peeling process. The first step to derive its expression is to write down the mechanical energy $\mathcal{E}$ of the system comprising the thin film and the loading system. It is composed of the two following terms
\begin{equation}
\mathcal{E} = \iint_\mathcal{D} \mathcal{E}_\mathrm{el}({\bf \nabla u}, {\bf\nabla w}, {\bf\nabla^2 w}) dx \, dy - \int_\mathcal{B} (F_x u_x + F_z w) dl
\label{Eq_1}
\end{equation}
where the elastic energy $\mathcal{E}_\mathrm{el}$ by unit surface stored in the debonded region of the thin film is a function of the gradient of the displacements and the curvatures of the thin film, as detailed in the following. The second term provides the work of the external force.

As peeling proceeds, both in-plane and out-of-plane displacements vary. As a result, we introduce their time dependence ${\bf u} = {\bf u}(t)$ and $w = w(t)$ as well as the one $\mathcal{F} = \mathcal{F}(t)$ of the peeling front. In the following, we calculate the {\it variations} of mechanical energy that provides both the equilibrium conditions applying on the displacement field and the elastic energy release rate along the peeling front. Varying Eq.~\eqref{Eq_1} with respect to time, we obtain the variations of mechanical energy
\begin{equation}
\begin{array}{lcl}
\displaystyle \frac{d \mathcal{E}}{dt} & = & \displaystyle \iint_\mathcal{D} \left( \frac{\partial \mathcal{E}_\mathrm{el}} { \partial u_{p,q} } \dot{u}_{p,q} +  \frac{\partial \mathcal{E}_\mathrm{el}} { \partial w_{,q} }  \dot{w}_{,q} + \frac{ \partial \mathcal{E}_\mathrm{el}} { \partial w_{,ij} } \dot{w}_{,ij} \right) dx \, dy
\vspace{15pt} \\ 
	 & &  \displaystyle - \int_{\mathcal{B}} (F_\mathrm{x} \dot{u} + F_\mathrm{z} \dot{w} ) dl + \int_\mathcal{F} \mathcal{E}_\mathrm{el} \,  v \, dl
\end{array}
\end{equation}
where the Einstein notation $x_i y_i = \sum\limits_{i = 1}^{2} x_i y_i$ for the sommations is adopted. In the previous expression, $v$ represents the normal velocity of the adhesion front. The first term is integrated by parts so the variation of mechanical energy is decomposed in the following four terms
\begin{equation}
\begin{array}{lcl}
\displaystyle \frac{d \mathcal{E}}{dt} & = & \displaystyle \iint_\mathcal{D} \left( - \left(\frac{\partial \mathcal{E}_\mathrm{el}} {\partial  u_{p,q} }  \right)_{,q} \dot{u}_p + 
\left[ -
\left(  \frac{\partial \mathcal{E}_\mathrm{el}} { \partial w_{,q}} \right)_{,q} 
+ \left( \frac{\partial \mathcal{E}_\mathrm{el}} {\partial w_{,ij}} \right)_{,ij} \right] \dot{w} \right) dx \, dy
\vspace{15pt} \\
 & & \displaystyle +   \int_{\mathcal{B_\circ}} \left[ \frac{\partial \mathcal{E}_\mathrm{el}}{\partial u_{p,q} } \dot{u}_p n_q + \left(  \frac{\partial \mathcal{E}_\mathrm{el}}{\partial w_{,q}} n_q - \frac{\partial^2 \mathcal{E}_\mathrm{el}} {\partial w_{,ij} \partial x_j} n_j \right) \dot{w} + \frac{\partial \mathcal{E}_\mathrm{el}}{\partial w_{,ij} } \dot{w_{,i}} n_j \right] dl
\vspace{15pt} \\
 & & \displaystyle + \int_{\mathcal{B}} \left( \left[ ... \right] - F_x \dot{u} - F_z \dot{w} \right) dl+\int_{\mathcal{F}} \left(  \left[ ... \right] +   \mathcal{E}_\mathrm{el} \, v \right)  \, dl
\end{array}
\label{Eq_3}
\end{equation}
where {\bf n} is the normal vector to the boundary specified in each integral, namely $\mathcal{B}_\circ$, $\mathcal{B}$ or $\mathcal{F}$. We retrieve the very same integrand noted $[ ...]$ in the three boundary integrals.

We now specify this last expression to the particular plate model considered in this study, namely the non-linear plate theory of ~\cite{Foppl} and~Von~\cite{VonKarman}, for which the elastic energy per unit surface follows
\begin{equation}
 \mathcal{E}_\mathrm{el} =  \frac{1}{2} \, M_{ij} w_{,ij} +  \frac{h}{2} \,  \sigma_{ij} \epsilon_{ij}.
\label{Eq_Eel}
\end{equation}
$M_{ij}$ and  $\sigma_{ij}$ are the moments and the stress field, respectively. $M_{ij}$ that represents the internal bending moments, varies with the position $x$ in the plate. In the F\"oppl - Von Karman plate model, the strain is related to the displacements through the relations
\begin{equation}
\epsilon_{ij} =  \frac{1}{2} \, (u_{i,j} + u_{j,i}) +  \frac{1}{2} \, w_{,i} w_{,j} .
\label{Eq_10}
\end{equation}
We emphasize that the F\"oppl - Von Karman plate theory includes contributions from {\it bending} and {\it stretching}, both modes of deformation being a priori coupled.

The stress field and the moments are given by the expressions
\begin{equation}
\sigma_{ij} = \frac{1}{h} \frac{\partial  \mathcal{E}_\mathrm{el}}{\partial  u_{i,j}}  \quad \quad M_{ij} = \frac{\partial  \mathcal{E}_\mathrm{el}}{\partial w_{,ij}}
\label{Eq_4}
\end{equation}
that can be derived from the elastic energy and the strains given in  Eqs.~\eqref{Eq_Eel} and~\eqref{Eq_10}. In the following, the stress tensor is simply noted $\sigma({\bf x})$.

The previous expressions can now be used in Eq.~\eqref{Eq_3} to express the variations of mechanical energy as a function of the stresses and the moments as
\begin{equation}
\begin{array}{lcl}
\displaystyle \frac{d\mathcal{E}}{dt} & = & \displaystyle \iint_\mathcal{D} \left ( - h \sigma_{ij,j} \, \dot{u}_i + \left [ - \left( h\sigma_{ji} w_{,j} \right)_{,i} + M_{ij,ij} \right ] \dot{w} \right) dx \, dy
\vspace{4pt} \\
 & & \displaystyle + \int_\mathcal{{B}_\circ} \left [ h \sigma_{ij} n_j  \, \dot{u}_i + (h \sigma_{ij} w_{,j} n_i - M_{ij,j} n_i) \, \dot{w} + M_{ij} \, \dot{w}_{,i} n_j \right ] dl
\vspace{4pt} \\
 & & \displaystyle + \int_{\mathcal{B}} \left [ (h \sigma_{ij} n_j - F_x) \, \dot{u}_i + (h \sigma_{ij} w_{,j} n_i - M_{ij,j} n_i - F_z ) \, \dot{w} + M_{ij} \, \dot{w}_{,i} n_j  \right ] dl
\vspace{4pt} \\
 & & \displaystyle + \int_{\mathcal{F}} \left [  \mathcal{E}_\mathrm{el} - h \sigma_{ij} n_j u_{i,q} n_q - h \sigma_{ij} w_{,j} n_i w_{,q} n_q + M_{ij,j} n_i w_{,q} n_q  -  M_{ij} n_j w_{,iq} n_q \right ]  v \, dl .
\end{array}
\label{Eq_Energy}
\end{equation}
The expression of the boundary integrals are simplified using the relationships $\{ \dot{u}_{p}  =  - v \, u_{p,q} n_q, \dot{w}  =  - v \, u_{,p} n_p, \dot{w}_{,p}  = - v \, w_{,pq} n_q  \}$ that derive from the continuity conditions $\{ w[c(t),t] = 0, u_{,p}[c(t),t] = 0, w_{,p}[c(t),t] = 0 \}$ of the displacement and the strain along the adhesion front. 

During peeling, the mechanical energy $\mathcal{E}$ {\it decreases} and the released energy is dissipated along the adhesion front $\mathcal{F}$. In Eq.~\eqref{Eq_Energy}, we identify the fourth integral along $\mathcal{F}$ as the {\it total} energy released that derives from the {\it local} elastic energy release rate $G$ through the relation $\displaystyle \frac{d\mathcal{E}}{dt} = - \int_\mathcal{F} G \, v  \, dl $. Owing to energy conservation, the three other integrals must be zero. As shown below, they provide the equilibrium conditions governing the stress field  $\sigma(\bf{x})$ and the adhesive shape $w(\bf{x})$ submitted to time-dependent boundary conditions on the peeling front.

Noticing that the displacements $\{ {\bf u} , w \}$ and their derivatives can be varied independently from each other, one obtains the following set of equilibrium equations
\begin{equation}
(\mathcal{D}) \left\{
\begin{array}{l}
\sigma_{ij,j} = 0
\vspace{2pt} \\
M_{ij,ij} - h \sigma_{ij} w_{,ij} = 0
\end{array}
\right.
\label{Eq_Int1}
\end{equation}
\begin{equation}
(\mathcal{B}_0) \left\{
\begin{array}{l}
\sigma_{ij} n_j = 0
\vspace{2pt} \\
M_{ij} n_j = 0
\vspace{2pt} \\
h \sigma_{ij} w_{,j} n_i - M_{ij,j} n_i = 0
\end{array}
\right.
\label{Eq_Int2}
\end{equation}
\begin{equation}
(\mathcal{B}) \left\{
\begin{array}{l}
h \sigma_{ij} n_j = F_x
\vspace{2pt} \\
F_x w_{,j} = F_z + M_{ij,j} n_i
\vspace{2pt} \\
M_{ij} n_j = 0.
\end{array}
\right.
\label{Eq_Int3}
\end{equation}

Combining the two expressions of Eq.~\eqref{Eq_Int1}, one retrieves the equilibrium equations of a F\"oppl - Von Karman plate
\begin{equation}
\left(M_{ij,i} - h \sigma_{ij} w_{,i} \right)_{,j} = 0
\label{Eq_Equi}
\end{equation}
that will be solved later using the boundary conditions~\eqref{Eq_Int2} and~\eqref{Eq_Int3} along the film boundary and the ones derived below applying along the adhesion front.

We now come back on the fourth integral in Eq.~\eqref{Eq_Energy} defined along the adesion front $\mathcal{F}$ that provides the mechanical energy flowing away from the adhesive and being dissipated in the process zone. Using the definition $\displaystyle \frac{d\mathcal{E}}{dt} = - \int_\mathcal{F} G \, v \, dl$ of the elastic energy release rate, we obtain $G = - \mathcal{E}_\mathrm{el} + h \sigma_{ij} n_j u_{i,q} n_q +  h \sigma_{ij} w_{,j} n_i w_{,q} n_q - M_{ij} w_{,iq} n_j n_q$.
This expression can be simplified using $\sigma_{ij} n_j u_{i,q} n_q = \sigma_{ij} n_j u_{i,q} n_q + (\sigma_{ij} t_j) (u_{i,q} t_q) = \sigma_{ij} u_{i,j} $
since $u_{i,p} t_p = 0$. Showing the same way that $\sigma_{ij} w_{,j} n_i w_{,q} n_q =\sigma_{ij} w_{,j} w_{,i}$ and $M_{ij} n_j w_{,iq} n_q = M_{ij} w_{,ij}$, one obtains $G = - \mathcal{E}_\mathrm{el} + \sigma_{ij} \epsilon_{ij}  + M_{ij} w_{,ij}$.  Recognizing the expression~\eqref{Eq_Eel} of the elastic energy, we obtain
\begin{equation}
G = \mathcal{E}_\mathrm{el} .
\label{Eq_Feq}
\end{equation}
Strikingly, the elastic energy release rate is set by the elastic energy by unit surface stored next to the adhesion front. Despite this seemingly local criterion, we will see later that the elastic energy release rate is a {\it non-local} function of the adhesion front deformations. The reason behind the emergence of non-local effects is the existence of {\it fields}, $w(\bf{x})$ and $\sigma(\bf{x}),$ that couple regions of the adhesion front far away from each other, so that the local energy density $\mathcal{E}_\mathrm{el}$ at a particular location along the front is a function of its deformation in a surrounding region.

Applying this expression to the case of an Euler-Bernoulli plate of bending stiffness $D = E h^3/(12(1-\nu^2))$, we predict $G = 1/2 D \, w_{,xx}^2$ that corresponds to the expression obtained by~\cite{Legrand2} taking into account the bending mode of deformation only.\footnote{The expression (17) of the elastic energy release rate in the article of~\cite{Legrand2} must be divided by two before comparing it with our findings as they consider a crack lying at the interface between {\it two} plates.}

In the following, we consider both modes of deformation, namely bending and stretching. Thus, we write the elastic energy $\mathcal{E}_\mathrm{el} $ of a F\"oppl - Von Karman plate as a function of the stress field and its curvatures (see {\it e.g.}~\cite{Audoly2}) to obtain
\begin{equation}
\begin{array}{l c c}
\displaystyle G & = & \displaystyle \frac{D}{2} \left[ ( w_{,xx} + w_{,yy})^2 - 2(1-\nu) (w_{,xx} w_{,yy} - w_{,xy}^2 ) \right]
\vspace{10pt}
\\
\displaystyle &  & + \displaystyle  \frac{h}{2E} \left[ (\sigma_{xx} + \sigma_{yy} )^2 - 2 (1+\nu) (\sigma_{xx} \sigma_{yy} - \sigma_{xy}^2 \right]
\label{Eq_General}
\end{array}
\end{equation}
where the bending stiffness $D = E h^3/(12(1-\nu^2))$ is set by the Young's modulus $E$ and the Poisson's ratio $\nu$ of the thin film material. We emphasize that in this expression, the stresses and the curvatures must be computed along the adhesion front. We notice here that the elastic energy release rate $G = G_\mathrm{b} + G_\mathrm{s}$ writes as the sum of two terms. The first contribution $G_\mathrm{b}$, proportional to the bending stiffness $D$, that scales as $\propto h^3$ with the film thickness, is emerging from the {\it bending} mode of deformation while the second contribution $G_\mathrm{s}$ that scales as $\propto h$ results from the {\it stretching} of the adhesive. Yet, it does not mean that both deformation modes can be treated independently from each other and do not interplay. Indeed, the curvatures and the stresses involved in Eq.~\eqref{Eq_General} are set by the equilibrium equations~\eqref{Eq_Equi} that do take into account the coupling between both modes.

\section{Homogeneous peeling}
\label{Sec2}
\begin{figure}[h]
\includegraphics[width=10.cm]{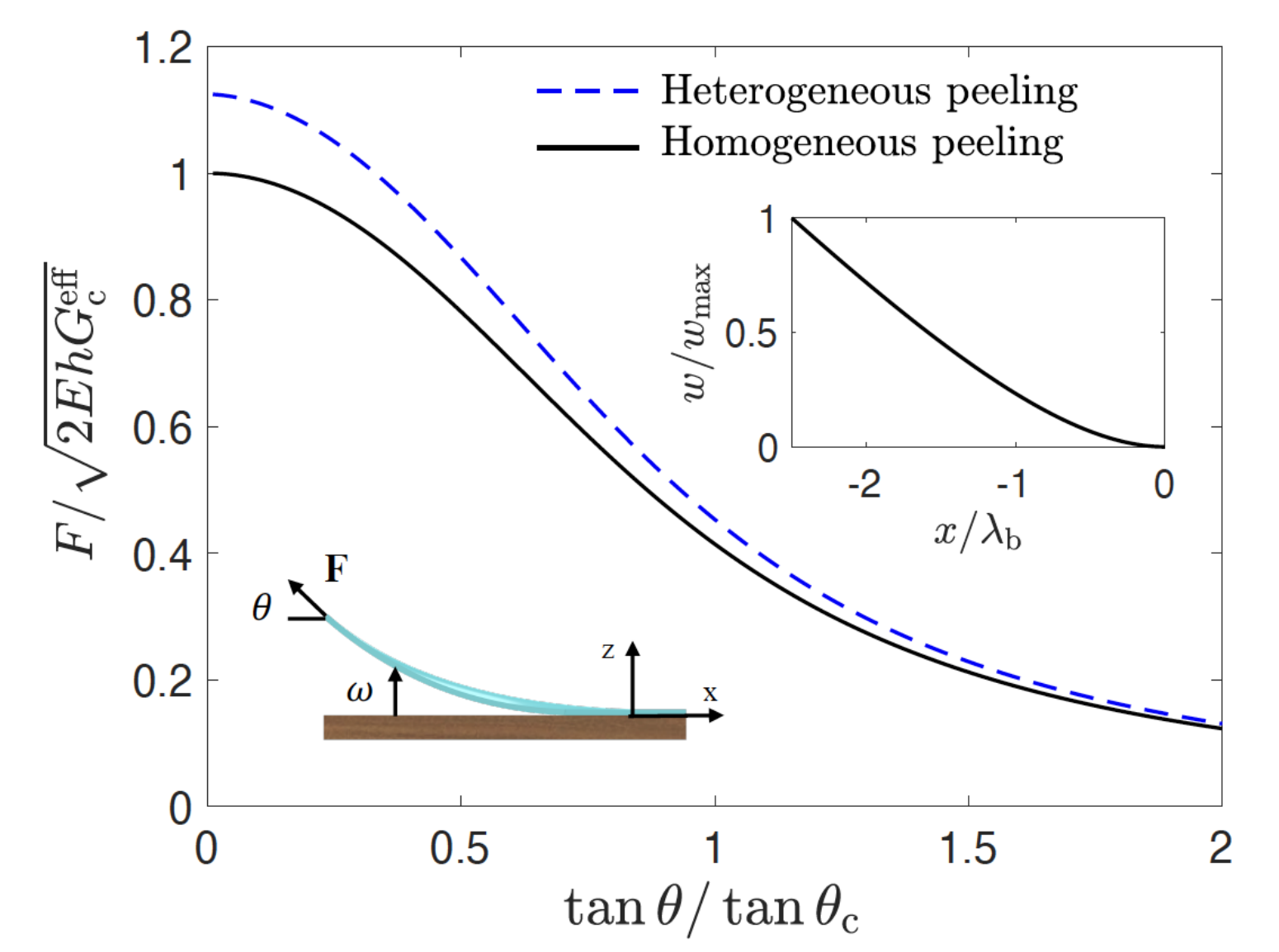}
\centering
\caption{Variations of the peel force with the peeling angle, as predicted by Eq.~\eqref{Eq_Straight}. The force is normalized by the maximum peel force $F = \sqrt{2 E h G_\mathrm{c}}$ obtained under shear for $\theta = 0$. $\theta_\mathrm{c}$ defined in Eq.~\eqref{Eq_thetac} corresponds to the critical peeling angle at the transition between stretching (for $\theta \ll \theta_\mathrm{c}$) and bending  (for $\theta \gg \theta_\mathrm{c}$) driven peeling. The dotted line shows the impact of random variations of adhesion energy of relative amplitude $\sigma = 50\%$, see Sec~\ref{Sec_Gceff} for more details. The inset shows the shape $w(x)$ of the adhesive as predicted from Eq.~\eqref{Eq_w} normalized by its position $w_\mathrm{max} = w(-L)$ at the application of the force. The position $x$ along the peeling direction is normalized by the bending length defined in Eq.~\eqref{Eq_lambda}, so that the adhesive shape is independent of the peeling angle.}
\label{Fig2}
\end{figure}

We explore the predictions of the proposed model under the hypothesis of {\it homogeneous} adhesion properties resulting in a {\it straight} adhesion front. Under some conditions discussed at the end of this section, the problem at hand becomes {\it invariant} along the adhesion front direction. As a result, the predicted elastic energy release rate can be compared with 2D peeling models such as those of~\cite{Rivlin} and~\cite{Kendall}. As stresses and displacements are functions of the peeling direction $x$ only, several simplifications are in order. The equilibrium equation~\eqref{Eq_Int1} of the stress field writes as $\displaystyle \frac{\partial \sigma_{xx}}{\partial x} =  \frac{\partial \sigma_{xy}}{\partial x} = 0$ that, after integration, gives $\sigma_{xx} = F/h$ and $\sigma_{yy} = \sigma_{xy} = 0$ using the boundary conditions~(\ref{Eq_Int2}) and~(\ref{Eq_Int3}). The second equilibrium condition~(\ref{Eq_Int1}) writes as $M_{xx,xx} - h \sigma_{xx} w_{,xx} = 0$ that can be integrated as $M_{xx} - h \sigma_{xx} w = \mathcal{C}_0 x + \mathcal{C}_1$ where $\mathcal{C}_0$ and $ \mathcal{C}_1$ are constants inferred from the boundary conditions. Using the expression~\eqref{Eq_4} of the moment, one finds $M_{xx} = \displaystyle D ( w_{,xx} + \nu \, w_{,yy}$), from which one obtains $w_{,xx} - w/\lambda_b^2 =  \mathcal{C}_2 x + \mathcal{C}_3$ where we define the bending length
\begin{equation}
\lambda_\mathrm{b} = \sqrt{D/F}.
\label{Eq_lambda}
\end{equation}
Applying the boundary conditions $\{ w(0) = 0, w_{,x}(0) = 0, w'(-L) = \tan\theta, w_{,x}(-L) = 0, \}$, we solve the previous differential equation and determine the constants $\mathcal{C}_2$ and $\mathcal{C}_3$, providing the out-of-plane displacement
\begin{equation}
w(x) = \lambda_\mathrm{b} \tan \theta \left[ e^{x/\lambda_\mathrm{b}} - \frac{x}{\lambda_\mathrm{b}} - 1 \right]
\label{Eq_w}
\end{equation}
of the thin film. This expression valid for $x<0$ goes beyond available peeling models that consider one or the other mode of deformation. As shown in the inset of Fig.~\ref{Fig2}, it describes the {\it shape} of the adhesive during its peeling. It is particularly relevant for analyzing experimental peel tests (see {\it e.g.}~\cite{Barlett} for a recent review). We see that beyond the bending length $|x| \gg \lambda_\mathrm{b}$, the adhesive shape $w(x) \simeq - \tan \theta \, (x + \lambda_\mathrm{b})$ is essentially flat, inclined with an angle $\theta$ set by the inclination of the peel force.

Derivating twice the previous expression, we obtain the radius of curvature $R_\mathrm{b} = 1/w_{,xx}(0) = \lambda_\mathrm{b}/\tan\theta$ of the thin film at the adhesion front location. We note that for small peeling angles $\theta \rightarrow 0$, the radius of curvature becomes very large. Using Euler-Bernoulli plate model as in~\cite{Legrand} and \cite{Xia} who considered bending deformations only, the elastic energy rate that scales as $G \propto 1/R_\mathrm{b}^2$ as reminded below would go to zero, leading to the divergence of the peel force for $\theta \rightarrow 0$. We will see that low angle peeling can be captured once the stretching mode is also taken into account. The elastic energy release rate of a F\"oppl - Von Karman thin film with a straight front is derived from the general expression~(\ref{Eq_General}), leading to
\begin{equation}
G =  \frac{D}{2} w_{,xx}^2 +  \frac{h}{2 E} \sigma_{xx}^2.
\end{equation}
The first term $G_\mathrm{b} \propto  w_{,xx}^2$ that derives from the bending energy of the film vanishes when $\theta \rightarrow 0$. It is consistent with the expression computed by~\cite{Legrand2} who used the Euler-Bernoulli beam model. The second term $G_\mathrm{s} \propto \sigma_{xx}^2$ derives from the stretching of the film. Using the previous expressions of the curvature and the stress, one obtains
\begin{equation}
G = G_\mathrm{b} + G_\mathrm{s} \quad \mathrm{with} \quad
\left\{
\begin{array}{l}
G_\mathrm{b} = \displaystyle \frac{1}{2}~ F \,  \tan^2 \theta 
\vspace{10pt} \\
G_\mathrm{s} = \displaystyle \frac{1}{2} \frac{F^2}{E h} .
\end{array} \right.
\label{Eq_Straight}
\end{equation}

In the limit of small peeling angles $\theta \ll 1$, one retrieves the linearized expression $G \simeq F \theta^2/2 + F^2/(2 E h)$ of the elastic energy release rate $G = F (1-\cos\theta) + F^2/(2 E h)$ derived by~\cite{Kendall} for extensible thin films. The comparison between both expressions illustrates that our approach applies for small peeling angles only, a limitation that results from the F\"oppl - Von Karman plate theory that remains valid for small inclinations $w_{,x} \ll 1$, and thus small peeling angles. From the comparison between Eq.~\eqref{Eq_Straight} and~\cite{Rivlin}'s exact result $G = F (1-\cos \theta)$ for inextensible ($E \rightarrow \infty$) thin films, we infer that our model provides an accurate expression of the elastic energy release rate (and thus of the peel force) within $5~\%$ for a peel angle $\theta < 25^\circ$ and within $10~\%$ for a peel angle $\theta < 35^\circ$. Yet, it appears that replacing $(1/2) \tan^2 \theta \rightarrow (1 - \cos \theta)$ that both behave as $\simeq \theta^2/2$ at small peeling angles provides an expression of the driving force valid for any peeling angle.

Using the equation of motion $(G-G_\mathrm{c})\, v \geq 0$ of quasi-static cracks~\citep{Griffith, Rice5} where $G_\mathrm{c}$ is the adhesion energy at the substrate-film interface, we obtain the peel force
\begin{equation}
F = \sqrt{ 2 E \, h  G_\mathrm{c}} \left ( \frac{\tan \theta } {\tan \theta_\mathrm{c} } \right )^2 \left ( \sqrt{1 + \left ( \frac{\tan \theta_\mathrm{c}}{\tan \theta} \right )^4} - 1 \right ) \quad \Rightarrow \quad
\left\{
\begin{array}{lcl}
F & \simbelow & \displaystyle \sqrt{2 E h G_\mathrm{c}}
\vspace{10pt} \\
F & \simabove & \displaystyle 2 \, G_\mathrm{c}/\tan^2 \theta
\end{array} \right.
\label{Eq_Fpeel}
\end{equation}
shown in Fig.~\ref{Fig2}. The critical peeling angle
\begin{equation}
\theta_\mathrm{c} = \arctan \left( \sqrt[4]{\frac{8 G_\mathrm{c}}{E h}} \right )
\label{Eq_thetac}
\end{equation}
separates a stretching driven regime for $\theta \ll \theta_\mathrm{c}$ from a bending driven regime for $\theta \gg \theta_\mathrm{c}$. The presence of a fourth square root in the expression~\eqref{Eq_thetac} of $\theta_\mathrm{c}$ renders the critical peeling angle {\it weakly} dependent on the adhesive properties, so that, for typical values of adhesive fracture energy, Young's modules and thickness, it ranges within $5^\circ \lesssim \theta_\mathrm{c} \lesssim 30^\circ$.

Our model is compatible with the expression $F = \sqrt{2 E h G_\mathrm{c}}$ of the peel force derived by~\cite{Kendall} for extensible thin films for $\theta \rightarrow 0$ and with the peel force $F = G_\mathrm{c}/(1-\cos \theta)$ derived by~\cite{Rivlin} for inextensible thin films for $\theta \gg \theta_\mathrm{c}$. Equation~\eqref{Eq_Fpeel} therefore constitutes a unified expression of the peel force valid in both regimes and at the transition between them. It remains accurate for small peeling angles, typically smaller than $35^\circ$. Yet, it can be generalized to any peeling angle by replacing $(1/2) \tan^2 \theta$ by $(1 - \cos \theta)$ as discussed earlier.  We then postulate that all the expressions derived in our study, including Eq.~\eqref{Eq_w}, derived under the assumption $\theta \ll 1$, remain valid at large peeling angles by changing  $(\tan^2 \theta)/2 \rightarrow 1 - \cos \theta$ .

We now would like to discuss an important assumption made in the previous analysis. Thin films under tension clamped to their extremity may {\it buckle} and form wrinkles parallel to the tensile loading direction, as reported by~\cite{Cerda}. This occurs because the clamped boundary prevents the plate to contract laterally, resulting, at some distance from the edge, to a lateral {\it compressive} stress, as shown by~\cite{Friedl}. During peeling, thin film adhesives also are clamped at the front location. Can such a buckling instability take place during the peeling of adhesives ? And if it occurs, would it impact their peeling behavior ?

The critical tensile force at which buckling takes place has been derived by~\cite{Jacques}. Through the comparison with the maximum peeling force occurring for $\theta \ll \theta_\mathrm{c}$, we infer the condition $L_\mathrm{c}/h \lesssim \pi/(C \,\tan \theta_\mathrm{c})$ for peeling to take place {\it without} buckling.\footnote{Comparing the critical tensile force $ F_\mathrm{b} = D \, \left ( \frac{2 \pi}{C L_\mathrm{c}} \right )^2$ derived by~\cite{Jacques} at which buckling takes place with the maximum peeling force $F = \sqrt{2 E h G_\mathrm{c}}$ provided in Eq.~\eqref{Eq_Fpeel}  for low peeling angles $\theta \ll \theta_\mathrm{c}$ leads to the condition  $F < F_\mathrm{b} \Leftrightarrow L_\mathrm{c}/h < \sqrt{2/[3(1-\nu^2)]} \, \pi/(C \tan \theta_\mathrm{c}) \simeq \pi/(C \tan \theta_\mathrm{c})$ for peeling to take place prior to buckling.} The extent $L_\mathrm{c}$ along $x$ of the region under lateral compressive stress is of the order of the adhesive width $W$ while the stress bi-axiality $C = \sigma_{yy}/\sigma_{xx} \simeq 5.10^{-3}$ prior to buckling is obtained numerically. The critical peeing angle $\theta_\mathrm{c}$ being smaller than 30° for most practical situations, we obtain the condition $W/h \lesssim 1000$ on the adhesive width which is generally satisfied.  Yet, what would be the impact of buckling if we were using adhesives of very large width ? As we will show in the next section, the presence of a bended region of size $\lambda_\mathrm{b}$ close to the adhesion front as shown in Fig.~\ref{Fig2} prevents wrinkles to extend along the peeling direction. We then expect wrinkles that would form due to buckling to be confined in the flat region $|x| \gg \lambda_\mathrm{b}$ of the adhesive, far from the adhesion front, thus preventing any interaction with the wrinkles studied in the following that results from the deformation of the adhesion front localized in its vicinity. This implies that both types of wrinkles might co-exist without interacting with each other, suggesting that buckling might have a limited effect on the peeling behavior of homogeneous and heterogeneous adhesives.

\section{Heterogeneous peeling}
\label{Sec3}
\subsection{Governing equations}
\label{Sec_G}
We now consider a {\it slightly} perturbed adhesion front $c(y) = c^{(0)} + \delta c(y)$ with $ \delta c(y) \ll c^{(0)}$ where $c$ is the total crack length.\footnote{With this definition, the crack length $c^{(0)}$ for a homogeneous adhesive is equal to the length $L$ of the debonded region of the adhesive.} The geometrical perturbations $\delta c$ may result from heterogeneous adhesion properties at the film-substrate interface. To solve the governing equations, we introduce the Airy potential $\phi({\bf x})$ (see {\it e.g.} \cite{Landeau}) related to the stress field through the expressions $\{ \sigma_{xx} = \phi_{,yy}, \sigma_{xy} = \phi_{,xy}, \sigma_{yy} = \phi_{,xx} \}$. We then decompose the out-of-plane displacement, the Airy potential, and the elastic energy release rate in two contributions
\begin{equation}
\left\{
\begin{array}{lcl}
w(\bold{x}) & = & w^{(0)}(x) + \delta w(\bold{x})
\vspace{10pt} \\
\phi(\bold{x}) & = & \phi^{(0)}(y) + \delta  \phi(\bold{x})
\vspace{10pt} \\
G(y) & = & G^{(0)} + \delta G(y).
\end{array} \right.
\label{Eq_Dvt}
\end{equation} 
The zeroth-order contributions $w^{(0)}$, $\sigma_{ij}^{(0)}$, $\phi^{(0)}$ and $G^{(0)}$ are solutions of the problem with a {\it straight} adhesion front studied in the previous section while the first-order contributions $\delta w$, $\delta \sigma_{ij}$, $\delta \phi$ and $\delta G$ result from the front perturbations and vanish when $\delta c \rightarrow 0$.

The stress field can be replaced by the Airy function in the general expression~(\ref{Eq_General}) of the elastic energy rate. Using the decomposition~\eqref{Eq_Dvt}, one then obtains the following expression
\begin{equation}
\delta G = D \frac{\partial^2 w^{(0)}} {\partial x^2} \left( \frac{\partial^2 \delta w} {\partial x^2} - \nu \frac{\partial^2 \delta w}{\partial y^2}  \right) + \frac{h}{E} \frac{\partial^2 \phi^{(0)}}{\partial y^2} \left( \frac{\partial^2 \delta \phi}{\partial y^2} - \nu  \frac{\partial^2 \delta \phi}{\partial x^2}  \right)
\end{equation}
that results from the dependence of $w^{(0)}$ with $x$ only and the one of $\phi^{(0)}$ with $y$ only. Using the zeroth order expression~\eqref{Eq_w} of the displacement and the stress derived in Section~\ref{Sec2}, one obtains
\begin{equation}
\delta G = \delta G_\mathrm{b} + \delta G_\mathrm{s} \quad \mathrm{with} \quad
\left\{
\begin{array}{lcl}
\delta G_\mathrm{b}(y) & = & \displaystyle F \lambda_b \,\tan \theta  \left(\frac{\partial^2 \delta w} {\partial x^2} - \nu \frac{\partial^2 \delta w}{\partial y^2}  \right) 
\vspace{10pt} \\
\delta G_\mathrm{s}(y) & = & \displaystyle \frac{F}{E} \,   \left( \frac{\partial^2 \delta \phi}{\partial y^2} - \nu  \frac{\partial^2 \delta \phi}{\partial x^2} \right).
\end{array} \right.
\label{Eq_Gbs}
\end{equation}
We recall that all the second derivatives involved in these equations must be applied along the adhesion front, in $u=0$;

In order to facilitate the resolution of the bulk equations~\eqref{Eq_Equi}, we introduce the dimensionless variables
\begin{equation}
\left\{
\begin{array}{lcl}
u = x/  \lambda_\mathrm{b} \quad & \mathrm{and} & v = y/\lambda_\mathrm{b} 
\vspace{10pt} \\
f = \delta w/\lambda_\mathrm{b} \quad & \mathrm{and} & g = \delta \phi/(\lambda_\mathrm{b}^2 E) .
 \end{array} \right.
\label{Eq_NoDim}
\end{equation}
The equilibrium equations on the dimensionless perturbations of the displacement and the stress thus write as
\begin{equation}
\left\{
\begin{array}{l}
\displaystyle f_{,uuuu} + f_{,vvvv} +2 f_{,uuvv} -  \, f_{,uu} = \frac{Eh}{F}  \tan \theta \, e^{u} g_{,vv} 
\vspace{10pt} \\
\displaystyle g_{,uuuu} + g_{,vvvv} + 2 g_{,uuvv} = - \tan \theta \, e^{u} f_{,vv}.
\end{array} \right.
\label{Eq_Bulk}
\end{equation} 
The four boundary conditions applying on the adhesion front in $u = 0$ and the four ones derived from Eq.~\eqref{Eq_Int3} at the extremity $u_\mathrm{c} = -c^{(0)}/\lambda_\mathrm{b}$ of the adhesive write as $\{ f(0, v) = 0 , f_{,u}(0, v) = - \tan \theta \, \delta c/\lambda_\mathrm{b}, (2 + \nu) g_{,uvv} (0,v) + g_{,uuu}(0,v) = - \delta c_{,vv} \, F \cos \theta/(E h \lambda_\mathrm{b} ) , g_{,uu}(0, v) - \nu g_{,vv}(0, v) = 0, f_{,u}(u_\mathrm{c},v) = 0, f_{,uu}(u_\mathrm{c},v) = 0 ,  g_{,uu}(u_\mathrm{c},v) = 0 , g_{,vv}(u_\mathrm{c},v) = 0 \}$.\footnote{The first two boundary conditions stem from the vanishing slope $w_{,x}[\delta c(y)] = 0$ of the adhesive at the adhesion front implying $\delta w_{,x}(0) = f_{,u}(0, v) = - w^{(0)}_{,xx}(0) \, \delta c(y)$ where the curvature $w^{(0)}_{,xx}(0) = \tan \theta/\lambda_\mathrm{b}$ of the unperturbed adhesive close to the adhesion front derives from Eq.~\eqref{Eq_w}. Similarly, the continuity condition $w[\delta c(y)] = 0$ of the displacement at the adhesion front leads to $\delta w(0) = - w^{(0)}_{,x}(0) \, \delta c(y) = 0 \Rightarrow f(0,v) = 0$. Note that this last relation implies $f_{,vv}(0,v) = 0$ used subsequently to simplify the expression of the bending energy release rate perturbations in Eq.~\eqref{Eq_fg}. A similar approach is used to derive the boundary conditions applying on the perturbations $g$ of the Airy function in $u=0$.} Using these boundary conditions, the expression~\eqref{Eq_Gbs} of the elastic energy release rate simplifies as
\begin{equation}
\left\{
\begin{array}{l}
\displaystyle \delta G_\mathrm{b}(v) = F \, \tan \theta  \, (f_{,uu}(0,v) - \nu f_{,vv}(0,v)) = F \, \tan \theta   \, f_{,uu}(0,v)
\vspace{10pt} \\
\displaystyle \delta G_\mathrm{s}(v) = F   (g_{,vv}(0,v) - \nu g_{,uu}(0,v)) =  F \,  (1 - \nu^2 )\, g_{,vv}(0,v) .
\end{array} \right.
\label{Eq_fg}
\end{equation}

\subsection{Decoupling of the displacement field and the stress field for small wavelength perturbations}
\label{Sec_Decouple}
Solving the coupled equations~(\ref{Eq_Bulk})  is challenging due to the presence of exponentials. Yet, for small wavelength perturbations of the adhesion front, the displacement perturbations {\it decouple} from the stress perturbations. As shown in~\ref{Appendix}, the right hand side of both equations can be neglected for $\lambda \ll 2 \pi \lambda_\mathrm{b}$ where the bending length $\lambda_\mathrm{b}$ defined in Eq.~\eqref{Eq_lambda} sets the radius of curvature of the adhesive. In Fourier space, this results in
\begin{equation}
\left\{
\begin{array}{l}
\displaystyle \hat f_{,uuuu} - (2 q^2 + 1) \hat f_{,uu} + q^4 \hat f = 0
\vspace{10pt} \\
\displaystyle \hat g_{,uuuu} - 2 q^2 \hat g_{,uu} + q^4 \hat g = 0.
\end{array} \right.
\label{Eq_Fourier}
\end{equation}
We introduced here the Fourier transform $\hat f(u,v)$ of the dimensionless displacement and the one $\hat g(u,v)$ of the dimensionless Airy function
\begin{equation}
\hat f(u,q) = \frac{1}{2\pi} \int f(u,q) e^{-i q v} dv \quad \mathrm{and} \quad \hat g(u,q) = \frac{1}{2\pi} \int g(u,q) e^{-i q v} dv.
\end{equation}
where $q$ is the dimensionless Fourier mode defined as $q = k \lambda_b$ while $k$ is the wave number along the $y$-axis. In the following, we will also use the Fourier transform $\hat{\delta c}$ of the front perturbations and the ones of the stress field perturbations 
\begin{equation}
\delta \hat{c}(q) = \frac{1}{2\pi} \int  \delta c(v) e^{-i q v} dv
\end{equation}
that will be used subsequently.

\subsection{Stretching contribution to the elastic energy release rate}
\label{Sec_Gs}
We are now in position to compute the distribution of elastic energy release rate along a slightly perturbed adhesion front. We focus first on the stretching mode of deformation described by the stress field $\sigma({\bf{x}})$ that derives from the Airy function $\phi({\bf{x}})$. As shown in the following, this deformation mode dominates at low peeling angle and hence sets the elastic energy release rate for $\theta \ll \theta_\mathrm{c}$.

\subsubsection{Airy function}
\label{Sec_Airy}
We start by calculating the perturbations of the Airy function described by the second equation of the system~\eqref{Eq_Fourier}. Its general expression is $\hat{g}(u,q) = [A(q) u + B(q)] e^{qu} + [C(q) u + D(q)] e^{-qu}$ that simplifies as $\hat g(u,q) = [\tilde{A}(q) u + \tilde{B}(q)] e^{|q| u}$ as the stress perturbations vanish far from the adhesion front, for $u \ll -1$. The boundary conditions $\{ - (2 + \nu) q^2 \hat g_{,u}(0,q) + \hat g_{,uuu}(0,q) = q^2 \frac{F}{E h} \frac{\hat{\delta c}(q)}{\lambda_\mathrm{b}}, \hat g_{,uu}(0,q) + \nu q^2 \hat g(0,q) = 0,\hat g_{,uu}(u_\mathrm{c},q) = 0 ,  \hat g(u_\mathrm{c},q) = 0\}$ provided in Section~\ref{Sec_G} have been rewritten here in Fourier space

The third and fourth boundary conditions are satisfied, irrespective of the constants $\tilde{A}(q)$ and $\tilde{B}(q)$, as we assume that the adhesive length $u_\mathrm{c}$ is much larger than the wavelength of the front perturbations. The first and second boundary conditions provide the constants $\tilde{A}(q)$ and $\tilde{B}(q)$, leading to
\begin{equation}
\hat g(u,q) = \frac{1}{3 - \nu} \frac{F}{E h} \left(\frac{2}{|q|(1+\nu)} - u \right) e^{|q| u} \frac{\hat{\delta c(q)}}{\lambda_\mathrm{b}}
\label{Eq_Airy}
\end{equation}

\subsubsection{Stress field}
The stress field perturbations $(\delta \sigma_{xx}$, $\delta \sigma_{yy})$ are derived from the Airy function perturbations from the relations $\{ \delta \sigma_{xx} = \frac {\partial^2 \delta \phi}{\partial y^2} = E \frac {\partial^2 g}{\partial v^2} , \delta \sigma_{yy} = \frac {\partial^2 \delta \phi}{\partial x^2} = E  \frac {\partial^2 g}{\partial u^2} \}$ in direct space and as $\{ \hat{\delta \sigma_{xx}}(u,q) = - E q^2 \hat g(u,q),  \hat{\delta \sigma_{yy}}(u,q) = E \hat g_{uu} \}$ in Fourier's space. $\hat{\delta \sigma_{xx}}(u,q)$ and $\hat{\delta \sigma_{yy}}(u,q)$ are the Fourier transforms of the stress field perturbations that then follow
\begin{equation}
\left\{
\begin{array}{l}
\hat{\delta \sigma_{xx}}(x,k) =  \displaystyle - \frac{1}{3 - \nu} \frac{F}{h} \left( \frac{2}{1+\nu} - |k| x \right) |k| \, \hat{\delta c}(k) \, e^{|k| x}
\vspace{10pt} \\
\hat{\delta \sigma_{yy}}(x,k) = \displaystyle - \frac{1}{3 - \nu} \frac{F}{h} \left( \frac{2 \nu}{1+\nu} + |k| x \right) |k| \, \hat{\delta c}(k) \, e^{|k| x}.
\end{array} \right.
\label{Eq25}
\end{equation}
Note that the stress field perturbations resulting from a adhesion front perturbation of wavelength $\lambda$ vanish at some distance $|x| \gg \lambda$ from the adhesion front, an observation in line with Saint-Venant's principle.

\subsubsection{Elastic energy release rate}
Using the expression~\eqref{Eq_Airy} of the Airy function perturbations in the general expression~\eqref{Eq_fg} of the elastic energy release, we obtain the stretching contribution
\begin{equation}
\delta \hat G_\mathrm{s}(k) = - \frac{2(1-\nu)}{3-\nu} \frac{F^2}{E h}  |k| \, \hat{\delta c}(k).
\label{Eq_Gs}
\end{equation}
Using the expression~\eqref{Eq_Straight} of the elastic energy release rate $G^{(0)}$ for a straigth adhesion front leads to 
\begin{equation}
 \frac{\delta \hat G_\mathrm{s}(k)}{G_\mathrm{s}^{(0)}} = - \frac{4(1-\nu)}{3-\nu} \,  |k| \, \delta \hat{c}(k).
\end{equation}

\subsection{Bending contribution to the elastic energy release rate}
\label{Sec_Gb}
We now focus on the perturbations of the displacement field that give rise to the bending contribution of the elastic energy release rate. It is governed by the first equation of the system~\eqref{Eq_Fourier} from which one obtains $\hat f(u,q) = A e^{q_+ u } + B e^{q_- u } + C e^{ - q_+ u } + D e^{ - q_- u }$ with $q_{\pm} = \sqrt{ q^2 + 1 /2 \pm \sqrt{q^2  + 1/4} }$. As the perturbation $\hat f(u,q)$ vanishes far from the adhesion front for $u \rightarrow -\infty$, one obtains $C=0$ and $D=0$. Using the two other boundary conditions on the adhesion front, namely $\hat f(0,q) = 0$ and $\hat f_{,u}(0,q) = - \tan \theta \, \delta \hat{c}(q) / \lambda_\mathrm{b}$, one obtains the normalized displacement field 
\begin{equation}
\hat f(u,q) = - \tan \theta \, \frac{e^{q_+ u} - e^{q_- u}}{q_+-q_-} \,  \frac{\delta \hat{c}(q)}{\lambda_b} .
\label{Eq_fsol}
\end{equation}
Derivating twice this expression and inserting it in Eq.~\eqref{Eq_fg} provides the perturbations of elastic energy release rate associated with bending
\begin{equation}
\delta \hat G_\mathrm{b}(q) = - F\, \tan^2 \theta  \,  \frac{ \delta \hat{c}(q)}{\lambda_b}(q_+ + q_-).
\label{Eq_Gbfin}
\end{equation}
Considering the small wavelength limit $\lambda \ll 2\pi \lambda_\mathrm{b} \Leftrightarrow q \gg 1 $, we obtain the bending contribution to the elastic energy release rate 
\begin{equation}
\delta \hat G_\mathrm{b}(k) = - 2 F\, \tan^2 \theta \, |k| \, \delta \hat{c}(k).
\label{Eq_Gb}
\end{equation}
Using the expression~\eqref{Eq_Straight} of the driving force for a straight adhesion front leads to
\begin{equation}
\frac{\delta \hat G_\mathrm{b}(k)}{G^{(0)}_\mathrm{b}} = - 4 \, |k|  \, \delta \hat c(k).
\label{Eq_Gb1}
\end{equation}
The very same expression has been derived by~\cite{Legrand2} who considered a crack lying at the interface between two elastic plates. This is consistent with the assumptions made in their model. First, using Love-Kirchhoff plate theory, they considered bending as the only deformation mode of the plate. In addition, they limited their calculations to small deflections of the plate that amounts to assume that its radius of curvature \lambdab is very large with respect to the front perturbations. Interestingly, the driving force perturbations along a adhesion front is exactly four times larger than the one along a perturbed crack front in a three-dimensional elastic medium~\citep{Rice4}, implying that adhesion fronts are four times {\it stiffer} than crack fronts.

It then turns out that under the assumptions of the small wavelength perturbations $\lambda \ll 2 \pi \lambda_\mathrm{b}$, both stretching and bending deformation modes give rise to long-range interactions that opposes to front perturbations. We will see later that depending on the peeling angle, one contribution prevails over the other one? We will discuss the implications on the peeling response of adhesives in Section~\ref{Sec4}.

What happens when the perturbation wavelength approaches $\lambda_\mathrm{b}$ ? As $\lambda \rightarrow \lambda_\mathrm{b}$, the coupling between the stress and the out-of-plane displacement builds up, and other terms $\delta G \propto - \delta c/\lambda_\mathrm{b}$ in the expression of the elastic energy release rate takes over (see~\ref{Appendix}). These terms that act as linear {\it local} restoring forces, {\it cut} the long-range elastic interactions: Two regions along the front separated by a distance $\delta y \gg \lambda_\mathrm{b}$ do not interact with each other. In practice, perturbations of wavelength larger than \lambdab are damped abnormally fast with respect to modes of smaller wavelength, as shown in Section~\ref{Sec4}. We now compute the out-of-plane displacement of the adhesive that will shed light on the origin of the long-range interactions and their cut-off at the length scale \lambdab.

\subsection{Wrinkles of perturbed adhesives}
\label{Sec_Wrinkle}
The wrinkles visible in Fig.~\ref{Fig1} are central features of the peeling behavior of heterogeneous adhesives. We investigate below their origin and the consequence on the adhesion front elasticity. Front deformations give rise to perturbations of the {\it shape} of the adhesive. These wrinkles extend along the peeling direction and ultimately explains the non-local elasticity of the adhesion front, as discussed below.

Considering the small wavelength approximation $\lambda \ll 2 \pi \lambda_\mathrm{b}$, the adhesive shape perturbations given by Eq.~\eqref{Eq_fsol} write as
\begin{equation}
\displaystyle \delta \hat w(x,k) = \displaystyle 2 \tan \theta  \, \sinh \left ( \frac{ |x| }{2 \lambda_\mathrm{b}} \right ) e^{|k| x}  \, \delta \hat{c}(k) .
\end{equation}

To obtain the displacement in the direct space, we consider sinusoidal perturbations $\delta c(y) = \delta c_\circ \sin(2 \pi y/\lambda)$ of the adhesion front of wavelength $\lambda$ and amplitude $\delta c_\circ$, giving rise to
\begin{equation}
\displaystyle \delta w(x,y)  = 2 \, \delta c_\circ \tan \theta \, \sinh \left ( \frac{ |x| }{2 \lambda_\mathrm{b}} \right ) \cos \left( \frac{2 \pi}{\lambda} y  \right)  e^{\frac{2 \pi x}{\lambda}}.
\end{equation}
The wavelength of the wrinkles that form on the adhesive are set by the wavelength $\lambda$ of the adhesion front. Their length along the peeling direction are also set by $\lambda$, as they vanish exponentially fast for $|x| \gg \lambda$. Their amplitude is proportional to the front perturbation amplitude $\delta c_\circ$. Yet, it decreases as the peeling angle decreases, a behavior that is expected as out-of-plane perturbations of the adhesive vanish in the limit $\theta \rightarrow 0$.

In \ref{Appendix}, we show that front perturbations of wavelength $\lambda \gg \lambda_\mathrm{b}$ do not give rise to wrinkles of length $\lambda$. Instead, their extension along the peeling direction is bounded to the bending length $\lambda_\mathrm{b}$. Elastic plates being reluctant to bend in two perpendicular directions at the same location, the overall bended shape of the adhesive prevents wrinkles to extend beyond their radius of curvature $\lambda_\mathrm{b}$. By hindering the development of wrinkles, this mechanism {\it cuts} the long-range interactions along the adhesive front.

This finding provides insights on the origin of the non-local elasticity of adhesion fronts. Front perturbations and thin film deformations are nested, so that a unidimensional perturbation of size $\lambda$ along the front impacts the adhesive over a bi-dimensional region of size $\lambda^2$. The energetic cost associated with such a bending deformation of the adhesive scales as $\delta E_\mathrm{b} \propto D \left(\delta c/\lambda \right)^2$, a behavior that explains the scaling of the bending energy release rate perturbations $\delta G_\mathrm{b} \propto - G_\mathrm{b}^{(0)} \delta c/\lambda$ derived from Eq.~\eqref{Eq_Gb1} for $\lambda \ll \lambda_\mathrm{b}$. In absence of characteristic length scales, the stiffness of the adhesion front is set by the wavelength of the front perturbations, a mechanism further discussed in Sec.~\ref{Sec_Stability}.

On the contrary, for front perturbations of size $\lambda \gg \lambda_\mathrm{b}$, long-range elasticity breaks down. The wrinkles that result from the front perturbations cannot extend beyond $|x| \gtrsim \lambda_\mathrm{b}$, suggesting that the unbounded development of perturbations pf the adhesive shape is key for non-local interactions to settle down.


Simply speaking, adhesion front elasticity reflects the reluctance of adhesives to bend and to stretch. Their non-local nature reflects a more subtle feature : The scale free interaction between the adhesion front and the field coupled to it,  here the  adhesive shape $\omega(\bf{x})$. Such a scale free interplay between a field and its boundary appears as a common feature of a wide variety of interfaces displaying long-range elasticty, such as crack fronts~\citep{Rice4}, wetting fronts~\citep{Joanny}, ferromagnetic domain walls~\citep{Zapperi2} or martensitic phase boundaries~\citep{Dondl}.



\section{Implications on the peeling behavior of adhesives}
\label{Sec4}
\subsection{From stretching to bending driven long-range elasticity}
\label{Sec_thetac}
Considering geometrical perturbations of the adhesion front of wavelength $\lambda \ll 2 \pi \lambda_\mathrm{b}$, we showed that the driving force perturbations $\delta G(y) = \delta G_\mathrm{b}(y) + \delta G_\mathrm{s}(y)$ decomposes into two contributions~\eqref{Eq_Gb} and~\eqref{Eq_Gs} associated with the bending energy and the stretching energy of the deformed adhesive that follow
\begin{equation}
\delta G(y) = \left (2 \, F \, \tan^2\theta  + \frac{2(1-\nu)}{3 - \nu}  \frac{F^2}{E h} \right ) \frac{1}{\pi} \mathrm{PV} \int_\infty^\infty \frac{\delta c(y') - \delta c(y)}{ (y'-y)^2 }  \, dy'
\label{Eq_Gtot} 
\end{equation}
where PV refers to the principal value of the integral. Irrespective of the peeling regime, the adhesion front elasticity is long-range, a feature that is discussed in Section~\ref{Sec_Stability}. Yet, the {\it stiffness} of the adhesion front, defined as the amplitude of the non-local front elasticity, {\it varies} with the peeling angle. Using the expression~\eqref{Eq_Straight} of the elastic energy release rate along a straight adhesion front, we obtain
\begin{equation}
\frac{\delta G(y)}{G^{(0)}} = \frac{\alpha(\theta)}{\pi} \mathrm{PV} \int_\infty^\infty \frac{\delta c(y') - \delta c(y)}{ (y'-y)^2} \, dy' \quad  \Leftrightarrow \quad \frac{\delta \hat G(k)}{G^{(0)}} = - \alpha(\theta) \, |k| \, \delta \hat c(k)
\label{Eq_Final}
\end{equation}
where the stiffness $\alpha(\theta)$ varies with the peeling angle as
\begin{equation}
\alpha(\theta) = \frac{4}{3-\nu} \left (1 - \nu + \frac{4}{1 + \sqrt{1 + \left ( \frac{\tan \theta_\mathrm{c}}{\tan \theta} \right)^4 }} \right ).
\end{equation}
\begin{figure}[h]
\includegraphics[width=10.cm]{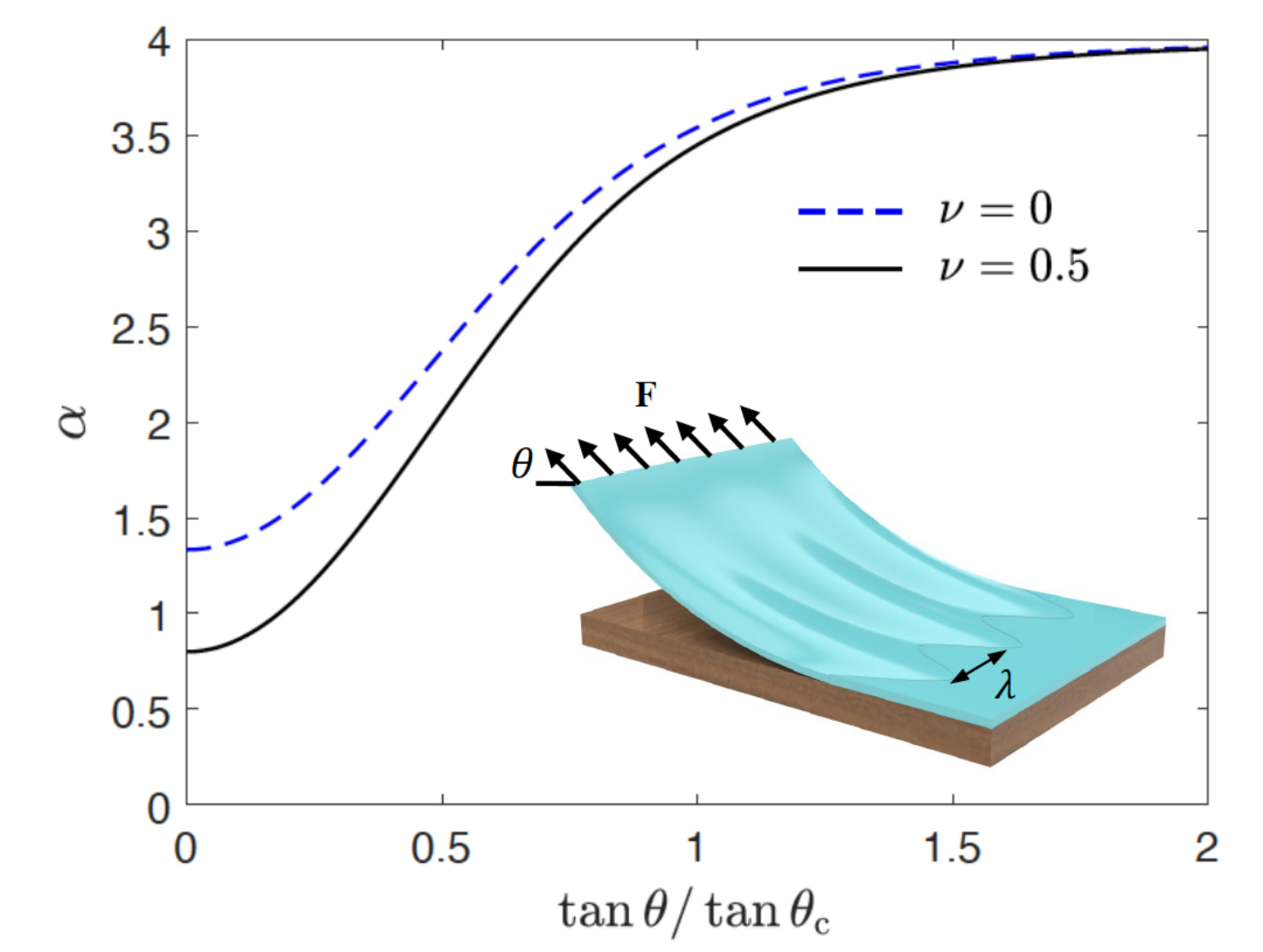}
\centering
\caption{Stiffness $\alpha(\theta)$ of the adhesion front $\delta \hat G(k) = - 4 \alpha(\theta) \, |k|  \, \delta \hat c(k)$ as a fonction of the peeling angle $\theta$ for the Poisson's ratios $\nu = 0$ and $\nu = 1/2$ where $k = 2\pi / \lambda$ are the deformation modes of the front with wavelength $\lambda$. The critical peeling angle $\theta_\mathrm{c}$ defined in Eq. \eqref{Eq_thetac} separates the peeling regime driven by the stretching of the adhesive at low angles from a regime driven by the bending of the adhesive at large angles for which the adhesion front is three to five times stiffer.}
\label{Fig_alpha}
\end{figure}
The stiffness is shown as a function of the peeling angle in Fig.~\ref{Fig_alpha} for $\nu = 0$ and $\nu = 1/2$. We evidence the transition from stretching dominated peeling for $\theta \ll \theta_\mathrm{c}$ to bending driven peeling for $\theta \gg \theta_\mathrm{c}$. In the bending regime, the stiffness $\alpha$ is equal to four, thus retrieving a non-local elasticity $\delta \hat{G} = - 4~G^{(0)} |k| \delta \hat c$ four times stronger than for cracks. For $\theta \ll \theta_\mathrm{c}$, when peeling is driven by the stretching of the adhesive, the stiffness $\alpha  = (1-\nu)/(3-\nu)$ is between five (for $\nu = 1/2$) and three (for $\nu = 0$) times smaller. Interestingly, we retrieve exactly the unit stiffness $\delta \hat{G} = - G^{(0)} |k| \delta \hat c$ of crack fronts for $\nu = 1/3$. The consequences of such a peeling angle dependent stiffness on the stability of the adhesion front, and more generally, on the peeling behavior of adhesives, are now investigated.

\subsection{Stability analysis of adhesion fronts}
\label{Sec_Stability}
We study below the stability of peeling fronts with respect to geometrical perturbations. Considering sinusoidal perturbations
\begin{equation}
\delta c(y) = \delta c_\circ \cos \left (\frac{2 \pi}{\lambda} y \right )
\label{Eq_Sin}
\end{equation}
of wavelength $\lambda$, one obtains the following distribution of elastic energy release rate
\begin{equation}
\frac{\delta G(y)}{G^{(0)}} = - 2 \pi \alpha(\theta) \frac{\delta c_\circ }{\lambda} \cos \left (\frac{2 \pi}{\lambda} y \right ).
\label{Eq_Sin}
\end{equation}
It appears that the front is {\it stable} with respect to geometrical perturbations. Indeed, the positions $y = n \, \lambda$ with $n \in \mathbb{Z}$ where the front is in advance correspond to the positions where the elastic energy release rate is smaller, implying that the sinusoidal perturbations will vanish. At low peeling angle, despite the absence of wrinkles, the in-plane deformations of the adhesive alone are sufficient to oppose to the front deformation, and thus ensure its stability. We also note from Eq.~\eqref{Eq_Sin} that the smaller the wavelength, the larger the restoring force, illustrating a basic feature of the long-range elasticity governing adhesion fronts.

Before exploring how the peeling angle dependent elasticity of adhesion fronts impacts their dynamical properties, we would like to discuss the fingering instability observed during the detachment of adhesives~\citep{Ghatak}. It turns out that our approach does not account for such an instability as it predicts that adhesion fronts are stable, even under shear. Yet, it reveals that fingering must take place at scales smaller than the wavelengths $\lambda \gg h$ of the stable modes analyzed in our model. This conclusion is in line with the observations made by~\cite{Ghatak4} who reported that the wavelength of the fingers are of the same order as the film thickness. It implies that plate's theory is unsuitable to study this phenomenon, and that 3D elasticity might be more adapted.

\subsection{Relaxation dynamics of adhesion fronts}
\label{Sec_Relax}
We now study how adhesion fronts recover a straight configuration after escaping from tough impurities. We are interested in understanding how the {\it relaxation time} is impacted by the peeling angle dependent front elasticity. We first derive the equation of motion of the adhesion front. We introduce a rate-dependent adhesion energy $G_\mathrm{c}(v)$ that, in most adhesives, depend on the front speed $v$ (see for example~\cite{Maugis2}). Assuming that the local front speed is strictly positive at any time step, the equation of motion of the adhesion front can be obtained from the balance $G = G_\mathrm{c}$ between the elastic energy released and the energy dissipated along the front.

Introducing the mean front speed $v_\mathrm{m} = \dot{c}_\circ$, the equation of motion of a straight front at zeroth order writes as $G^{(0)} = G_\mathrm{c}(v_\mathrm{m})$. At the first order in the front perturbations, it follows
\begin{equation}
\delta v/v_\circ = \delta G/G^{(0)} - \delta G_\mathrm{c}/\overline{G}_\mathrm{c}.
\label{Eq_Motion}
\end{equation}
where the material dependent speed $v_\circ = (dG_\mathrm{c}/dv)/G_\mathrm{c}(v_\mathrm{m})$ emerges from the rate-dependency of the adhesion energy~\citep{Chopin5}. Here, $\overline{G}_\mathrm{c}$ provides the average adhesion energy of the interface at the mean front speed $v_\mathrm{m}$ while $\delta G_\mathrm{c}(y,x) = G_\mathrm{c}(y,x) - \overline{G}_\mathrm{c}$ describes the spatial variations of adhesion energy. In the following, we note $\delta g_\mathrm{c} = (G_\mathrm{c} - \overline{G}_\mathrm{c})/\overline{G}_\mathrm{c}$ the relative variations of adhesion energy.
 
 Introducing the expression~\eqref{Eq_Final} of the driving force perturbations, we obtain the equation of motion in Fourier space
\begin{equation}
\delta \dot{c}/v_\circ = - \alpha(\theta)|k| \delta \hat c - \delta \hat{g}_\mathrm{c}.
\end{equation}
 Considering first the evolution of the adhesion front from an initial sinusoidal configuration $\delta c(y,t=0) = - \delta c_\circ [1 + \cos(2 \pi y/\lambda)]$ in a homogeneous interface $g_\mathrm{c} = 0$, Eq.~\eqref{Eq_Motion} predicts the following relaxation dynamics
\begin{equation}
\delta c(y, t) = - \delta c_\circ [1 + \cos(2 \pi y/\lambda)] e^{-t/\tau} \quad \mathrm{where} \quad \tau = \frac{\lambda}{2\pi \alpha(\theta) v_\circ}.
\end{equation}
We retrieve the main result of Section~\ref{Sec_Stability}: The adhesion front relaxes exponentially fast to a straight configuration, illustrating the stability of the peeling front with respect to geometrical perturbations. The relaxation time increases linearly with the perturbation wavelength, a feature that stems from the long-range nature of the front elasticity. We also observe that the peel angle impacts the relaxation time, a feature specific to adhesion fronts. The larger the peeling angle, the longer it takes for the front to recover a straight configuration. This behavior is reminiscent of the larger stiffness of the front for peeling angles $\theta \gg \theta_\mathrm{c}$. It also sheds light on the enhanced stability of the adhesion front for large peeling angles, when the bending mode of deformation of the adhesive drives the peeling process.

Considering now perturbations along the front larger than the bending length $\lambda_\mathrm{b}$, we show that the relaxation time $\tau \simeq \lambda_\mathrm{b}/v_\circ$ saturates and does not scale anymore with $\lambda$. This implies that perturbations with wavelength beyond the long-range elastic regime are damped abnormally fast with respect to the other modes, reflecting here the impact of the cut-off of the interactions derived in~\ref{Appendix}. In practice, this effect manifests on the power-spectrum of the adhesion front roughness in presence of disorder that shows a scale invariant behavior up to the cut-off length $\lambda_\mathrm{b}$~\citep{Chopin3}.

\subsection{Effective peeling resistance of disordered interfaces}
\label{Sec_Gceff}
We now study the impact of the peeling angle dependent front elasticity on the peeling resistance. In absence of adhesion heterogeneities, the peeling front remains straight, and the role of the front elasticity is rather minor, limited to ensuring the front stability. But in presence of adhesion heterogeneities, it starts to play an important role as it impacts the peel strength. For disordered interfaces, the {\it effective} adhesion energy - the apparent crack growth resistance that must be considered to predict the peel strength - is given by
\begin{equation}
G_\mathrm{c}^\mathrm{eff} \simeq \bar{G}_\mathrm{c} \left [ 1 + \sigma^2/\alpha(\theta) \right ]
\end{equation}
where $\sigma = \langle \delta g_\mathrm{c}^2(y,x) \rangle_{y,x}^{1/2} $ is the relative strength of the adhesion energy disorder~\citep{Demery,Lebihain}. As a result, the smaller the stiffness of the adhesion front, the larger the impact of the disorder on the effective adhesion energy. Using the relation~\eqref{Eq_Fpeel} between peel strength and adhesion energy, we obtain the two following asymptotic behaviors
\begin{equation}
\left\{
\begin{array}{lcl}
F & \simbelow & \displaystyle F_\mathrm{hom} \left [  1 + \sigma^2/(2 \alpha(\theta)) \right ] \simeq F_\mathrm{hom}  \left [  1 +  \frac{3-\nu}{2(1-\nu)} \, \frac{\sigma^2}{4} \right ] 
\vspace{10pt} \\
F & \simabove & \displaystyle F_\mathrm{hom} \left [ 1 + \sigma^2/\alpha(\theta) \right ]   \simeq F_\mathrm{hom}  \left [  1 + \frac{\sigma^2}{4}  \right ] 
\end{array} \right.
\label{Eq_Fpeelhet}
\end{equation}
where $F_\mathrm{hom}$ is the peel force for a homogeneous interface of adhesion energy $\bar{G}_\mathrm{c}$. The predicted peel force for a disorder of strength $\sigma = 50~\%$ is shown as a function of the peeling angle and compared with the homogeneous peel strength $F_\mathrm{hom}$ in Fig.~\ref{Fig2}. For the case considered with $\nu = 1/3$, the relative increase of peel force resulting from the disorder is twice larger at low peeling angle. It represents an increase of the peel strength of the adhesive of about $12~\%$ and  $6~\%$ at low and large peeling angle, respectively.

\section{Conclusions and perspectives}
\label{Sec5}
We analyzed theoretically the peeling behavior of elastic adhesives. Adopting a fracture mechanics framework and describing thin film adhesives as F{\"o}ppl-Von Karman plates, we derived the governing equation of the adhesion front from which the peel strength was inferred. The originality of our work is to consider the two main modes of deformation of adhesives, namely their stretching and their bending. We show that the peeling response crucially depends on the competition between this both modes. At large peeling angles, the bending mode takes over, giving rise to a peel force that scales linearly with the adhesion energy and decreases with the peeling angle, in line with~\cite{Rivlin}'s classical result. For small peeling angles, the stretching mode dominates. The peel strength tends toward a constant value that varies as the square root of the adhesion energy, the adhesive Young's modulus and its thickness, recovering \cite{Kendall}'s predictions. The critical peeling angle $\theta_\mathrm{c} = \arctan(\sqrt[4]{8G_\mathrm{c}/Eh})$ derived from our model separates both regimes. The shape of the adhesive predicted by our model is shown to be universal and independent of the peeling angle, once normalized by the bending length $\lambda_\mathrm{b} = \sqrt{D/F}$.

As soon as heterogeneities are introduced at the interface between the adhesive and the substrate, the adhesion front deforms. The front elasticity then governs both its geometry, its dynamics and impact the peel strength. To study these effects, we derive the adhesion front elasticity. We show that the adhesion elasticity is long-range, as for crack fronts, irrespectively of the peeling angle. Yet, the strength of this non-local elasticity varies with $\theta$. As for homogeneous peeling, we evidence two regimes separated by $\theta_\mathrm{c}$: At large peeling angles, the adhesion front is four times stiffer than crack fronts. This behavior is reminiscent of the wrinkles visible in Fig.~\ref{Fig1} that form on the adhesives as a result of the front deformations. This result is in line with the previous studies of~\cite{Legrand2} and~\cite{Xia3} who considered the bending mode of deformation only. At low peeling angles, the front elasticity results from the stretching of the adhesive and depends on the Poisson's ratio. Its stiffness is of the same order as crack fronts, thus much lower than for large peeling angles.

The consequence of such a peeling angle dependent front elasticity is finally investigated. Even though the stability of the peeling front with respect to geometrical perturbations is ensured by the non-local elasticity irrespective of $\theta$, we show that the enhanced stiffness of the adhesion front at large angles provides an enhanced stability of the peeling front. This feature is evidenced by the relaxation time of the adhesion front that is about four times smaller in this regime than at low peeling angle. As the peeling angle tunes the ability of the peel front to deform, it also impacts the peeling behavior in presence of adhesion heterogeneities. We show that the increase of peeling resistance resulting from the collective pinning of the front in presence of a disordered distribution of adhesion energy is about twice larger at small peeling angles.

Overall, this study highlights the central role played the {\it elastic deformations} of adhesives on their peeling behavior. The proposed framework provides tools to {\it engineer} the interplay between the front deformation and the peeling driving force. In the spirit of the recent works of~\cite{Xia4} and~\cite{Subbarayan}, appropriately designed heterogeneities can then impact the driving force and thus the peeling resistance of the adhesive. One interesting concept emerging from this work is that front elasticity is not immutable. It can vary with the loading conditions, and, to some extent, be selected to tune the interactions between heterogeneities and the peeling front, opening new opportunities for the design of programmable adhesives.

\vspace{10pt} 
\noindent {\bf Acknowledgement}
The author would like to dedicate this work to his former mentor, Pr Kaushik Bhattacharya.

\appendix
\section{Governing equation for small wavelength perturbations}
\label{Appendix}
We justify here why the non-linear system~\eqref{Eq_Bulk} of coupled differential equations applying on the displacement field and the Airy function can be decoupled for small wavelength perturbations $\lambda \ll 2 \pi \lambda_\mathrm{b}$. In Fourier space, the coupled system of equation writes as
\begin{equation}
\left\{
\begin{array}{l}
\displaystyle \hat f_{,uuuu} - (2 q^2 + \cos \theta) \hat f_{,uu} + q^4 \hat f = \frac{Eh}{F}  \tan \theta \, e^{u} q^2 \hat g 
\vspace{10pt} \\
\displaystyle \hat g_{,uuuu} - 2 q^2 \hat g_{,uu} + q^4 \hat g = - \tan \theta \, e^{u} q^2 \hat f.
\end{array} \right.
\label{Eq_Fourier2}
\end{equation}
We focus on the second equation. The term on the right-hand side is noted $T_1$ while the term on the left hand side is noted $T_0$. From Eq.~\eqref{Eq_fsol}, we can approximate the right hand side by $T_1 \simeq q^2 \tan^2 \theta /\sqrt{\cos \theta} \, e^{q u} \,  \delta \hat c / \lambda_\mathrm{b}$ while from Eq.~\eqref{Eq_Airy}, we infer that the dominant terms of  $T_0$ scale as $\propto q^4 e^{q u} \,  \delta \hat c / \lambda_\mathrm{b}$. In the limit of small wavelengths $q \gg 1$, we thus conclude that $T_1 \ll T_0$. The second term in the Eq.~\eqref{Eq_Fourier2} applying on $\hat g$ can thus be neglected. The resulting homogeneous linear differential equation has been solved in Sec.~\ref{Sec_Airy} and provides the expression~\eqref{Eq_Airy} of the Airy function.

We now insert the expression~\eqref{Eq_Airy} of the Airy function in Eq.~\eqref{Eq_Fourier2} providing a {\it non-homogeneous} linear differential equation applying on $\hat f$ only. Its solution consists of two terms: A general solution to the corresponding {\it homogeneous} linear differential equation that has been previously derived in Sec.~\ref{Sec_Gb} and a particular solution to the equation including the right hand side noted $\hat f^{(2)}$. As the bending contribution $\delta \hat G_\mathrm{b}$ of the elastic energy release rate derives from the displacement through the linear equation~\eqref{Eq_fg}, we infer that $\delta \hat G_\mathrm{b}$ consists of two terms: One first term $\delta \hat G^{(1)}$ that derives from the solution of the {\it homogeneous} equation already calculated in Sec.~\ref{Sec_Gb} provided in Eq.~\eqref{Eq_Gb1}, and an additional term noted $\delta \hat G^{(2)}$ that derives from the particular solution $\hat f^{(2)}$. After some lengthy calculation, we obtain the expression of this second term that writes as
\begin{equation}
\delta G^{(2)}(y) = - \frac{2}{3} \, \frac{2 - \nu}{3 - \nu} \, F   \tan^2 \theta \, \frac{\delta c(y)}{\lambda_\mathrm{b}}
\end{equation}
This term is negligible in comparison to the non-local term $\delta  G^{(1)}$  for $\lambda \ll 2 \pi \lambda_\mathrm{b}$, justifying that the right hand side of Eq.~\eqref{Eq_Fourier2} from which $\delta G^{(2)}$ emerges can be neglected. For $\lambda \simeq 2 \pi \lambda_\mathrm{b}$, both terms become of the same order, revealing that long-range elasticity is cut at length scales larger than $\lambda_\mathrm{b}$. It implies that the {\it coupling} between the stretching mode and the bending mode of deformation of the adhesive at play for large perturbation wavelengths gives rise to {\it local} front elasticity. This suggests that the interplay between competing modes of deformation may not promote non-local effects. On the contrary, for non-local interactions to emerge, a single mode of deformation must be prevail over the other ones.


\begin{thebibliography}{51}
\expandafter\ifx\csname natexlab\endcsname\relax\def\natexlab#1{#1}\fi
\providecommand{\url}[1]{\texttt{#1}}
\providecommand{\href}[2]{#2}
\providecommand{\path}[1]{#1}
\providecommand{\DOIprefix}{doi:}
\providecommand{\ArXivprefix}{arXiv:}
\providecommand{\URLprefix}{URL: }
\providecommand{\Pubmedprefix}{pmid:}
\providecommand{\doi}[1]{\href{http://dx.doi.org/#1}{\path{#1}}}
\providecommand{\Pubmed}[1]{\href{pmid:#1}{\path{#1}}}
\providecommand{\bibinfo}[2]{#2}
\ifx\xfnm\relax \def\xfnm[#1]{\unskip,\space#1}\fi
\bibitem[{Audoly and Pomeau(2010)}]{Audoly2}
\bibinfo{author}{Audoly, B.}, \bibinfo{author}{Pomeau, Y.},
  \bibinfo{year}{2010}.
\newblock \bibinfo{title}{Elasticity and geometry}.
\newblock \bibinfo{publisher}{Oxford University Press}.
\bibitem[{Barlett et~al.(2023)Barlett, Case, Kinloch and Dillard}]{Barlett}
\bibinfo{author}{Barlett, M.}, \bibinfo{author}{Case, S.W.},
  \bibinfo{author}{Kinloch, A.J.}, \bibinfo{author}{Dillard, D.A.},
  \bibinfo{year}{2023}.
\newblock \bibinfo{title}{Peel tests for quantifying adhesion and toughness: A
  review}.
\newblock \bibinfo{journal}{Prog. Mat. Sci.} \bibinfo{volume}{137},
  \bibinfo{pages}{101086}.
\bibitem[{bhattacharya(2003)}]{Bhattacharya2}
\bibinfo{author}{bhattacharya, K.}, \bibinfo{year}{2003}.
\newblock \bibinfo{title}{Microstructure of martensite: Why It forms and how It
  gives rise to the shape-memory effect}.
\newblock \bibinfo{publisher}{Oxford University Press}.
\bibitem[{Bonamy et~al.(2008)Bonamy, Santucci and Ponson}]{Bonamy5}
\bibinfo{author}{Bonamy, D.}, \bibinfo{author}{Santucci, S.},
  \bibinfo{author}{Ponson, L.}, \bibinfo{year}{2008}.
\newblock \bibinfo{title}{Crackling dynamics in material failure as the
  signature of a self-organized dynamic phase transition}.
\newblock \bibinfo{journal}{Phys. Rev. Lett.} \bibinfo{volume}{101},
  \bibinfo{pages}{045501}.
\bibitem[{Cerda et~al.(2002)Cerda, Ravi-Chandar and Mahadevan}]{Cerda}
\bibinfo{author}{Cerda, E.}, \bibinfo{author}{Ravi-Chandar, K.},
  \bibinfo{author}{Mahadevan, L.}, \bibinfo{year}{2002}.
\newblock \bibinfo{title}{Wrinkling of an elastic sheet under tension}.
\newblock \bibinfo{journal}{Nature} \bibinfo{volume}{419},
  \bibinfo{pages}{579}.
\bibitem[{Chopin et~al.(2018)Chopin, Bhaskar, Jog and Ponson}]{Chopin5}
\bibinfo{author}{Chopin, J.}, \bibinfo{author}{Bhaskar, A.},
  \bibinfo{author}{Jog, A.}, \bibinfo{author}{Ponson, L.},
  \bibinfo{year}{2018}.
\newblock \bibinfo{title}{Depinning dynamics of crack fronts}.
\newblock \bibinfo{journal}{Phys. Rev. Lett.} \bibinfo{volume}{121},
  \bibinfo{pages}{235501}.
\bibitem[{Chopin et~al.(2015)Chopin, Boudaoud and Adda-Bedia}]{Chopin3}
\bibinfo{author}{Chopin, J.}, \bibinfo{author}{Boudaoud, A.},
  \bibinfo{author}{Adda-Bedia, M.}, \bibinfo{year}{2015}.
\newblock \bibinfo{title}{Morphology and dynamics of a crack front propagating
  in a model disordered material}.
\newblock \bibinfo{journal}{J. Mech. Phys. Solids} \bibinfo{volume}{74},
  \bibinfo{pages}{38--48}.
\bibitem[{Combescot(2022)}]{Combescot}
\bibinfo{author}{Combescot, P.}, \bibinfo{year}{2022}.
\newblock \bibinfo{title}{Superconductivity}.
\newblock \bibinfo{publisher}{Cambridge University Press}.
\bibitem[{Dalmas et~al.(2009)Dalmas, Barthel and Vandembroucq}]{Dalmas2}
\bibinfo{author}{Dalmas, D.}, \bibinfo{author}{Barthel, E.},
  \bibinfo{author}{Vandembroucq, D.}, \bibinfo{year}{2009}.
\newblock \bibinfo{title}{Crack front pinning by design in planar heterogeneous
  interfaces}.
\newblock \bibinfo{journal}{J. Mech. Phys. Solids} \bibinfo{volume}{57},
  \bibinfo{pages}{446--457}.
\bibitem[{D\'emery et~al.(2014)D\'emery, Rosso and Ponson}]{Demery}
\bibinfo{author}{D\'emery, V.}, \bibinfo{author}{Rosso, A.},
  \bibinfo{author}{Ponson, L.}, \bibinfo{year}{2014}.
\newblock \bibinfo{title}{From microstructural features to effective toughness
  in disordered brittle solids}.
\newblock \bibinfo{journal}{EPL} \bibinfo{volume}{105}, \bibinfo{pages}{34003}.
\bibitem[{Dondl and Bhattacharya(2010)}]{Dondl}
\bibinfo{author}{Dondl, P.}, \bibinfo{author}{Bhattacharya, K.},
  \bibinfo{year}{2010}.
\newblock \bibinfo{title}{A sharp interface model for the propagation of
  martensitic phase boundaries}.
\newblock \bibinfo{journal}{Arch. Rat. Mech. An.} \bibinfo{volume}{197},
  \bibinfo{pages}{599--617}.
\bibitem[{F\"oppl(1907)}]{Foppl}
\bibinfo{author}{F\"oppl, A.}, \bibinfo{year}{1907}.
\newblock \bibinfo{title}{Vorlesungen \"uber Technische Mechanik}.
  volume~\bibinfo{volume}{5}.
\newblock \bibinfo{publisher}{B. G. Teubner}, \bibinfo{address}{Leipzig}.
\bibitem[{Fultz(2020)}]{Fultz}
\bibinfo{author}{Fultz, B.}, \bibinfo{year}{2020}.
\newblock \bibinfo{title}{Phase transitions in materials}.
\newblock \bibinfo{publisher}{Cambridge University Press}.
\bibitem[{Gao and Rice(1986)}]{Gao4}
\bibinfo{author}{Gao, H.}, \bibinfo{author}{Rice, J.R.}, \bibinfo{year}{1986}.
\newblock \bibinfo{title}{Shear stress intensity factors for a planar crack
  with a slightly curved front}.
\newblock \bibinfo{journal}{J. Appl. Mech.} \bibinfo{volume}{53},
  \bibinfo{pages}{774--778}.
\bibitem[{Gao and Rice(1987)}]{Gao3}
\bibinfo{author}{Gao, H.}, \bibinfo{author}{Rice, J.R.}, \bibinfo{year}{1987}.
\newblock \bibinfo{title}{Somewhat circular crack}.
\newblock \bibinfo{journal}{Int. J. Frac.} \bibinfo{volume}{33},
  \bibinfo{pages}{155--174}.
\bibitem[{Gao and Rice(1989)}]{Gao}
\bibinfo{author}{Gao, H.}, \bibinfo{author}{Rice, J.R.}, \bibinfo{year}{1989}.
\newblock \bibinfo{title}{A first-order perturbation analysis of crack trapping
  by arrays of obstacles}.
\newblock \bibinfo{journal}{J. Appl. Mech.} \bibinfo{volume}{56},
  \bibinfo{pages}{828--836}.
\bibitem[{Gennes(2002)}]{DeGennes2}
\bibinfo{author}{Gennes, P.G.D.}, \bibinfo{year}{2002}.
\newblock \bibinfo{title}{Gouttes, bulles, perles et ondes}.
\newblock \bibinfo{publisher}{Belin}.
\bibitem[{Ghatak and Chaudhury(2003)}]{Ghatak4}
\bibinfo{author}{Ghatak, A.}, \bibinfo{author}{Chaudhury, M.K.},
  \bibinfo{year}{2003}.
\newblock \bibinfo{title}{Adhesion-iiduced instability patterns in thin
  confined elastic film}.
\newblock \bibinfo{journal}{Langmuir} \bibinfo{volume}{19},
  \bibinfo{pages}{2621--2631}.
\bibitem[{Friedl et~al.(2000)Friedl, Rammerstorfer and Fischer}]{Friedl}
\bibinfo{author}{Friedl, N.}, \bibinfo{author}{Rammerstorfer, F.G.},
  \bibinfo{author}{Fischer, F.D.}, \bibinfo{year}{2000}.
\newblock \bibinfo{title}{Buckling of stretched strips}.
\newblock \bibinfo{journal}{Comp. Struc.} \bibinfo{volume}{78},
  \bibinfo{pages}{185--190}.
\bibitem[{Ghatak et~al.(2000)Ghatak, Chaudhury, Shenoy and Sharma}]{Ghatak}
\bibinfo{author}{Ghatak, A.}, \bibinfo{author}{Chaudhury, M.K.},
  \bibinfo{author}{Shenoy, V.}, \bibinfo{author}{Sharma, A.},
  \bibinfo{year}{2000}.
\newblock \bibinfo{title}{Meniscus instability in a thin elastic film}.
\newblock \bibinfo{journal}{Phys. Rev. Lett.} \bibinfo{volume}{85},
  \bibinfo{pages}{4329--4332}.
\bibitem[{Griffith(1920)}]{Griffith}
\bibinfo{author}{Griffith, A.A.}, \bibinfo{year}{1920}.
\newblock \bibinfo{title}{The phenomena of rupture and flow in solids}.
\newblock \bibinfo{journal}{Phil. Trans. Roy. Soc. Lond.}
  \bibinfo{volume}{A221}, \bibinfo{pages}{163--198}.
\bibitem[{Jacques and Potier-Ferry(2005)}]{Jacques}
\bibinfo{author}{Jacques, N.}, \bibinfo{author}{Potier-Ferry, M.},
  \bibinfo{year}{2005}.
\newblock \bibinfo{title}{On mode localisation in tensile plate buckling}.
\newblock \bibinfo{journal}{C. R. M\'ecanique} \bibinfo{volume}{333},
  \bibinfo{pages}{804--809}.
\bibitem[{Joanny and Gennes(1984)}]{Joanny}
\bibinfo{author}{Joanny, J.F.}, \bibinfo{author}{Gennes, P.G.D.},
  \bibinfo{year}{1984}.
\newblock \bibinfo{title}{A model for contact angle hysteresis}.
\newblock \bibinfo{journal}{J. Chem. Phys.} \bibinfo{volume}{81},
  \bibinfo{pages}{552}.
\bibitem[{Karman(1910)}]{VonKarman}
\bibinfo{author}{Karman, T.V.}, \bibinfo{year}{1910}.
\newblock \bibinfo{title}{Festigkeitsprobleme im maschinenbau}, in:
  \bibinfo{editor}{Teubner, B.} (Ed.), \bibinfo{booktitle}{Encyclop\"adie der
  Mathematischen Wissenschaften}, \bibinfo{address}{Leipzig}. p.
  \bibinfo{pages}{349}.
\bibitem[{Kendall(1973)}]{Kendall}
\bibinfo{author}{Kendall, K.}, \bibinfo{year}{1973}.
\newblock \bibinfo{title}{Thin film peeling-elastic term}.
\newblock \bibinfo{journal}{J. Phys. D} \bibinfo{volume}{8},
  \bibinfo{pages}{105--117}.
\bibitem[{Landeau and Lifchitz(1959)}]{Landeau}
\bibinfo{author}{Landeau, L.}, \bibinfo{author}{Lifchitz, E.},
  \bibinfo{year}{1959}.
\newblock \bibinfo{title}{Theory of elasticity}.
\newblock \bibinfo{publisher}{Pergamon Press}.
\bibitem[{Landeau and Lifshitz(1984)}]{Landeau2}
\bibinfo{author}{Landeau, L.D.}, \bibinfo{author}{Lifshitz, E.M.},
  \bibinfo{year}{1984}.
\newblock \bibinfo{title}{Electrodynamics of continuous media}.
\newblock \bibinfo{publisher}{Pergamon Press}.
\bibitem[{Lawn(1993)}]{Lawn}
\bibinfo{author}{Lawn, B.}, \bibinfo{year}{1993}.
\newblock \bibinfo{title}{Fracture of brittle solids}.
\newblock \bibinfo{publisher}{Cambridge University Press}.
\bibitem[{Lazarus(2011)}]{Lazarus2}
\bibinfo{author}{Lazarus, V.}, \bibinfo{year}{2011}.
\newblock \bibinfo{title}{Perturbation approaches of a planar crack in linear
  elastic fracture mechanics: a review}.
\newblock \bibinfo{journal}{J. Mech. Phys. Solids} \bibinfo{volume}{59},
  \bibinfo{pages}{121--144}.
\bibitem[{Lazarus and Leblond(1998)}]{Lazarus5}
\bibinfo{author}{Lazarus, V.}, \bibinfo{author}{Leblond, J.B.},
  \bibinfo{year}{1998}.
\newblock \bibinfo{title}{Three-dimensional crack-faceweigh tfunctions for the
  semi-infinite interface crack.i.variation of the stress intensity factors due
  to some small perturbation of the crack front}.
\newblock \bibinfo{journal}{J. Mech. Phys. Solids} \bibinfo{volume}{46},
  \bibinfo{pages}{4437--4455}.
\bibitem[{Lazarus and Leblond(2002)}]{Lazarus4}
\bibinfo{author}{Lazarus, V.}, \bibinfo{author}{Leblond, J.B.},
  \bibinfo{year}{2002}.
\newblock \bibinfo{title}{n-plane perturbation of the tunnel-crack under shear
  loading. ii: Determination of the fundamental kernel}.
\newblock \bibinfo{journal}{Int. J. Solids Struct.} \bibinfo{volume}{39},
  \bibinfo{pages}{4437--4455}.
\bibitem[{Lebihain et~al.(2021)Lebihain, Ponson, Kondo and Leblond}]{Lebihain}
\bibinfo{author}{Lebihain, M.}, \bibinfo{author}{Ponson, L.},
  \bibinfo{author}{Kondo, D.}, \bibinfo{author}{Leblond, J.B.},
  \bibinfo{year}{2021}.
\newblock \bibinfo{title}{Effective toughness of disordered brittle solids: A
  homogenization framework}.
\newblock \bibinfo{journal}{J. Mech. Phys. Solids} \bibinfo{volume}{153},
  \bibinfo{pages}{104463}.
\bibitem[{Leblond et~al.(2012)Leblond, Patinet, Frelat and Lazarus}]{Leblond2}
\bibinfo{author}{Leblond, J.B.}, \bibinfo{author}{Patinet, S.},
  \bibinfo{author}{Frelat, J.}, \bibinfo{author}{Lazarus, V.},
  \bibinfo{year}{2012}.
\newblock \bibinfo{title}{Second-order coplanar perturbation of a semi-infinite
  crack in an infinite body}.
\newblock \bibinfo{journal}{Eng. Frac. Mech.} \bibinfo{volume}{90},
  \bibinfo{pages}{129--142}.
\bibitem[{Leblond et~al.(1996)Leblond, S.~E and Perrin}]{Leblond6}
\bibinfo{author}{Leblond, J.B.}, \bibinfo{author}{S.~E, M.},
  \bibinfo{author}{Perrin, G.}, \bibinfo{year}{1996}.
\newblock \bibinfo{title}{The tensile tunnel-crack with a slightly wary front}.
\newblock \bibinfo{journal}{Int. J. Solids Struct.} \bibinfo{volume}{33},
  \bibinfo{pages}{1995--2022}.
\bibitem[{Legrand and Leblond(2010a)}]{Legrand}
\bibinfo{author}{Legrand, L.}, \bibinfo{author}{Leblond, J.B.},
  \bibinfo{year}{2010}a.
\newblock \bibinfo{title}{Evolution of the shape of the fronts of a pair of
  semi-infinite cracks during their coplanar coalescence}.
\newblock \bibinfo{journal}{ZAMM} .
\bibitem[{Legrand et~al.(2011)Legrand, Patinet, Leblond, Frelat, Lazarus and
  Vandembroucq}]{Legrand2}
\bibinfo{author}{Legrand, L.}, \bibinfo{author}{Patinet, S.},
  \bibinfo{author}{Leblond, J.B.}, \bibinfo{author}{Frelat, J.},
  \bibinfo{author}{Lazarus, V.}, \bibinfo{author}{Vandembroucq, D.},
  \bibinfo{year}{2011}.
\newblock \bibinfo{title}{Coplanar perturbation of a crack lying on the
  mid-plane of a plate}.
\newblock \bibinfo{journal}{Int. J. Frac.} \bibinfo{volume}{170},
  \bibinfo{pages}{67--82}.
\bibitem[{Legrand and Leblond(2010b)}]{Legrand3}
\bibinfo{author}{Legrand, N.}, \bibinfo{author}{Leblond, J.B.},
  \bibinfo{year}{2010}b.
\newblock \bibinfo{title}{In-plane perturbation of a system of two coplanar
  slit-cracks -- ii: Case of close inner crack fronts or distant outer ones}.
\newblock \bibinfo{journal}{Int. J. Solids Struct.} \bibinfo{volume}{47},
  \bibinfo{pages}{3504--3512}.
\bibitem[{Maugis and Barquins(1988)}]{Maugis2}
\bibinfo{author}{Maugis, D.}, \bibinfo{author}{Barquins, M.},
  \bibinfo{year}{1988}.
\newblock \bibinfo{title}{Stick-Slip and Peeling of Adhesive Tapes}.
\newblock \bibinfo{publisher}{Springer}, \bibinfo{address}{Dordrecht}.
\bibitem[{Patinet et~al.(2013a)Patinet, Alzate, Barthel, Dalmas, Vandembroucq
  and Lazarus}]{Patinet}
\bibinfo{author}{Patinet, S.}, \bibinfo{author}{Alzate, L.},
  \bibinfo{author}{Barthel, E.}, \bibinfo{author}{Dalmas, D.},
  \bibinfo{author}{Vandembroucq, D.}, \bibinfo{author}{Lazarus, V.},
  \bibinfo{year}{2013}a.
\newblock \bibinfo{title}{Finite size effects on crack front pinning at
  heterogeneous planar interfaces: Experimental, finite elements and
  perturbation approaches}.
\newblock \bibinfo{journal}{J. Mech. Phys. Solids} \bibinfo{volume}{61},
  \bibinfo{pages}{311--324}.
\bibitem[{Patinet et~al.(2013b)Patinet, Vandembroucq and Roux}]{Patinet2}
\bibinfo{author}{Patinet, S.}, \bibinfo{author}{Vandembroucq, D.},
  \bibinfo{author}{Roux, S.}, \bibinfo{year}{2013}b.
\newblock \bibinfo{title}{Quantitative prediction of effective toughness at
  random heterogeneous interfaces}.
\newblock \bibinfo{journal}{Phys. Rev. Lett.} \bibinfo{volume}{110},
  \bibinfo{pages}{165507}.
\bibitem[{Pindra et~al.(2010)Pindra, Lazarus and Leblond}]{Pindra2}
\bibinfo{author}{Pindra, N.}, \bibinfo{author}{Lazarus, V.},
  \bibinfo{author}{Leblond, J.B.}, \bibinfo{year}{2010}.
\newblock \bibinfo{title}{In-plane perturbation of a system of two coplanar
  slit-cracks -- i: Case of arbitrarily spaced crack fronts}.
\newblock \bibinfo{journal}{Int. J. Solids Struct.} \bibinfo{volume}{47},
  \bibinfo{pages}{3489--3503}.
\bibitem[{Ravi-Chandar et~al.(2023)Ravi-Chandar, Leblond, Ponson and
  Barthelat}]{Ponson23}
\bibinfo{author}{Ravi-Chandar, K.}, \bibinfo{author}{Leblond, J.B.},
  \bibinfo{author}{Ponson, L.}, \bibinfo{author}{Barthelat, F.},
  \bibinfo{year}{2023}.
\newblock \bibinfo{title}{Mechanics and physics of fracture: Multiscale
  Modeling of the Failure Behavior of Solids}.
\newblock \bibinfo{publisher}{Springer}.
\bibitem[{Rice(1978)}]{Rice5}
\bibinfo{author}{Rice, J.R.}, \bibinfo{year}{1978}.
\newblock \bibinfo{title}{Thermodynamics of the quasi-static growth of griffith
  cracks}.
\newblock \bibinfo{journal}{J. Mech. Phys. Solids} \bibinfo{volume}{26},
  \bibinfo{pages}{61--78}.
\bibitem[{Rice(1985)}]{Rice4}
\bibinfo{author}{Rice, J.R.}, \bibinfo{year}{1985}.
\newblock \bibinfo{title}{First-order variation in elastic fields due to
  variation in location of a planar crack front}.
\newblock \bibinfo{journal}{J. Appl. Mech.} \bibinfo{volume}{52},
  \bibinfo{pages}{571--579}.
\bibitem[{Rivlin(1944)}]{Rivlin}
\bibinfo{author}{Rivlin, R.S.}, \bibinfo{year}{1944}.
\newblock \bibinfo{title}{The effective work of adhesion}.
\newblock \bibinfo{journal}{Paint Technology} \bibinfo{volume}{9},
  \bibinfo{pages}{215}.
\bibitem[{Schmittbuhl et~al.(1995)Schmittbuhl, Roux, Vilotte and
  M{\aa}l{\o}y}]{Schmittbuhl4}
\bibinfo{author}{Schmittbuhl, J.}, \bibinfo{author}{Roux, S.},
  \bibinfo{author}{Vilotte, J.P.}, \bibinfo{author}{M{\aa}l{\o}y, K.J.},
  \bibinfo{year}{1995}.
\newblock \bibinfo{title}{Interfacial crack pinning: effect of nonlocal
  interactions}.
\newblock \bibinfo{journal}{Phys. Rev. Lett.} \bibinfo{volume}{74},
  \bibinfo{pages}{1787--1790}.
\bibitem[{Shen et~al.(2024)Shen, He, Chen, Yang and Jiang}]{Shen_J}
\bibinfo{author}{Shen, J.}, \bibinfo{author}{He, Z.}, \bibinfo{author}{Chen,
  H.}, \bibinfo{author}{Yang, Y.}, \bibinfo{author}{Jiang, H.},
  \bibinfo{year}{2024}.
\newblock \bibinfo{title}{Exploiting interfacial instability during peeling a
  flexible plate from elastic films}.
\newblock \bibinfo{journal}{J. Mech. Phys. Solids} \bibinfo{volume}{192},
  \bibinfo{pages}{105821}.
\bibitem[{Sosson et~al.(2005)Sosson, Chateauminois and Creton}]{Sosson}
\bibinfo{author}{Sosson, F.}, \bibinfo{author}{Chateauminois, A.},
  \bibinfo{author}{Creton, C.}, \bibinfo{year}{2005}.
\newblock \bibinfo{title}{Investigation of shear failure mechanisms of
  pressure-sensitive adhesives}.
\newblock \bibinfo{journal}{J. Pol. Sci.} \bibinfo{volume}{43},
  \bibinfo{pages}{3316--3330}.
\bibitem[{Vasoya et~al.(2016a)Vasoya, Lazarus and Ponson}]{Vasoya4}
\bibinfo{author}{Vasoya, M.}, \bibinfo{author}{Lazarus, V.},
  \bibinfo{author}{Ponson, L.}, \bibinfo{year}{2016}a.
\newblock \bibinfo{title}{Bridging micro to macroscale fracture properties in
  highly heterogeneous brittle solids: weak pinning versus fingering}.
\newblock \bibinfo{journal}{J. Mech. Phys. Solids} \bibinfo{volume}{95},
  \bibinfo{pages}{755--773}.
\bibitem[{Vasoya et~al.(2016b)Vasoya, Unni, Leblond, Lazarus and
  Ponson}]{Vasoya2}
\bibinfo{author}{Vasoya, M.}, \bibinfo{author}{Unni, A.B.},
  \bibinfo{author}{Leblond, J.B.}, \bibinfo{author}{Lazarus, V.},
  \bibinfo{author}{Ponson, L.}, \bibinfo{year}{2016}b.
\newblock \bibinfo{title}{A theoretical and experimental study of crack pinning
  by strong heterogeneities}.
\newblock \bibinfo{journal}{J. Mech. Phys. Solids} \bibinfo{volume}{89},
  \bibinfo{pages}{211--230}.
\bibitem[{Xia et~al.(2012)Xia, Ponson, Ravichandran and Bhattacharya}]{Xia}
\bibinfo{author}{Xia, S.}, \bibinfo{author}{Ponson, L.},
  \bibinfo{author}{Ravichandran, G.}, \bibinfo{author}{Bhattacharya, K.},
  \bibinfo{year}{2012}.
\newblock \bibinfo{title}{Toughening and asymmetry in peeling of heterogeneous
  adhesives}.
\newblock \bibinfo{journal}{Phys. Rev. Lett.} \bibinfo{volume}{108},
  \bibinfo{pages}{196101}.
\bibitem[{Xia et~al.(2013)Xia, Ponson, Ravichandran and Bhattacharya}]{Xia4}
\bibinfo{author}{Xia, S.}, \bibinfo{author}{Ponson, L.},
  \bibinfo{author}{Ravichandran, G.}, \bibinfo{author}{Bhattacharya, K.},
  \bibinfo{year}{2013}.
\newblock \bibinfo{title}{Adhesion of heterogeneous thin films---i: elastic
  heterogeneity}.
\newblock \bibinfo{journal}{J. Mech. Phys. Solids} \bibinfo{volume}{61},
  \bibinfo{pages}{838--851}.
\bibitem[{Xia et~al.(2015)Xia, Ponson, Ravichandran and Bhattacharya}]{Xia3}
\bibinfo{author}{Xia, S.}, \bibinfo{author}{Ponson, L.},
  \bibinfo{author}{Ravichandran, G.}, \bibinfo{author}{Bhattacharya, K.},
  \bibinfo{year}{2015}.
\newblock \bibinfo{title}{Adhesion of heterogeneous thin films: Ii. adhesive
  heterogeneity}.
\newblock \bibinfo{journal}{J. Mech. Phys. Solids} \bibinfo{volume}{83},
  \bibinfo{pages}{88--103}.
\bibitem[{Xu et~al.(2019)Xu, Singh, Pan and Subbarayan}]{Subbarayan}
\bibinfo{author}{Xu, Y.}, \bibinfo{author}{Singh, Y.}, \bibinfo{author}{Pan,
  C.}, \bibinfo{author}{Subbarayan, G.}, \bibinfo{year}{2019}.
\newblock \bibinfo{title}{Adhesive toughness and instability in bonded
  heterogeneous films}.
\newblock \bibinfo{journal}{Int. J. Solids Struct.} \bibinfo{volume}{169},
  \bibinfo{pages}{41--54}.
\bibitem[{Zapperi et~al.(1998)Zapperi, Cizeau, Durin and Stanley}]{Zapperi2}
\bibinfo{author}{Zapperi, S.}, \bibinfo{author}{Cizeau, P.},
  \bibinfo{author}{Durin, G.}, \bibinfo{author}{Stanley, H.E.},
  \bibinfo{year}{1998}.
\newblock \bibinfo{title}{Dynamics of ferromagnetic domain wall: avalanches,
  depinning transition and the barkhausen effect}.
\newblock \bibinfo{journal}{Phys. Rev. E} \bibinfo{volume}{58},
  \bibinfo{pages}{6353--6366}.

\end{thebibliography}

\end{document}